\documentclass[opre,nonblindrev]{informs3} 

\DoubleSpacedXI 

\usepackage{natbib}
 \bibpunct[, ]{(}{)}{,}{a}{}{,}%
 \def\bibfont{\small}%
 \def\bibhang{24pt}%
\TheoremsNumberedThrough     
\ECRepeatTheorems

\EquationsNumberedThrough    

\usepackage{bm}
\usepackage{enumerate}
\usepackage{nicefrac}
\usepackage{mathrsfs}
\usepackage{multirow}
\usepackage{footnotebackref}
\usepackage{capt-of}
\usepackage{hyperref}
\newcommand{\Long}[2]{#1} 
\newcommand{\new}[2]{#1} 
\renewcommand{\P}{{\mathcal P}}
\newcommand{\ulam}{{\underline \lambda}}
\newcommand{\x}{{\bf x}}
\newcommand{\xs}{ x}

\begin{document}


\RUNAUTHOR{Shiksha Singhal, Veeraruna Kavitha and Jayakrishnan Nair}

\RUNTITLE{On the ubiquity of duopolies in constant sum congestion games}

\TITLE{On the ubiquity of duopolies in constant sum congestion games\footnote{A preliminary version of this work was presented at the CDC conference~\cite{CDC}, which describes the problem formulation (Section~\ref{sec_model}) and presents some of the results in Section~\ref{stable_config} without proof. The present paper is a significant extension; it includes an impossibility result under classical notions of stability (Section~\ref{sec:classical}), an analysis of stable configurations under an extension of the Shapley value for partition form games (Section~\ref{stable_config}), a complete characterisation of stable configurations in heavy and light traffic regimes (Section~\ref{sec_two_partition}), and a comprehensive numerical case study (Section~\ref{sec_numerical}).
}}

\ARTICLEAUTHORS{%
\AUTHOR{Shiksha Singhal}
\AFF{IEOR, Indian Institute of Technology, Bombay, India \EMAIL{shiksha.singhal@iitb.ac.in}} 
\AUTHOR{Veeraruna Kavitha}
\AFF{IEOR, Indian Institute of Technology, Bombay, India \EMAIL{vkavitha@iitb.ac.in}}
\AUTHOR{Jayakrishnan Nair}
\AFF{EE, Indian Institute of Technology, Bombay, India \EMAIL{jayakrishnan.nair@ee.iitb.ac.in}}
} 

\ABSTRACT{%
\new{We analyse a coalition formation game between strategic service providers of a congestible service. The key novelty of our formulation is that it is a \emph{constant sum game}, i.e., the total payoff across all service providers (or coalitions of providers) is fixed, and dictated by the size of the market. The game thus captures the tension between resource pooling (to benefit from the resulting statistical economies of scale) and competition between coalitions over market share. In a departure from the prior literature on resource pooling for congestible services, we show that the grand coalition is in general not stable, once we allow for competition over market share.}{} In fact, under classical notions of stability (defined via blocking by \emph{any} coalition), we show that no partition is stable. This motivates us to introduce more restricted (and relevant) notions of blocking; \new{interestingly, we find that the stable configurations under these novel notions of stability are \emph{duopolies}, where the dominant coalition exploits its economies of scale to corner a disproportionate market share.}{} Furthermore, we completely characterise the stable duopolies in heavy and light traffic regimes.

\noindent
\textit{Subject Classification:} Games: Cooperative, Queues: Markovian 

\noindent
\textit{Area of Review: }Stochastic Models


}%



\maketitle

%

\newpage
\section{Introduction}

\new{Resource sharing is an efficient way of reducing congestion and
uncertainty in service industries. It
refers to an arrangement where service resources are pooled and used
jointly by a group (a.k.a., coalition) of providers, instead of each
provider operating alone using its own resources. Naturally, such a
coalition would be sustainable only if the participating providers
obtain higher payoffs than they would have obtained otherwise. The key
driver of coalition formation in congestion prone service systems is
the statistical economies of scale that emerge from the pooling of
service resources---this allows the coalition to offer a better
quality of service to its customers, and/or to attract more customers
to its service.

Not surprisingly, there is a considerable literature (for example, see~\cite{karsten} and the references therein) that analyses
resource pooling between independent providers of congestible services
via a cooperative game theoretic approach. In these papers, each
provider is modeled as a queueing system, with its own dedicated
customer base, that generates service requests according to a certain
arrival process. The payoff of each service provider is in turn
determined by the quality of service it is able to provide to its
(dedicated) customer base. In such a setting, the statistical
economies of scale from resource pooling typically drives the service
providers to pool all their servers together to form a \emph{grand
coalition}, which generates the greatest aggregate payoff across all
coalitional arrangements. Naturally, the resulting aggregate payoff
must be divided between the providers in a \emph{stable} manner, i.e.,
in such a way that no subset of providers has an incentive to `break
away' from the grand coalition. Such stable payoff allocations have
been demonstrated in a wide range of settings, including
single/multiple server environments, and loss/queue-based environments
(see~\cite{karsten,karsten2014}  and the references therein).

To summarize, the literature on coalition formation between providers
of congestible services suggests that a stable grand coalition would
emerge from the strategic interaction. However, a crucial aspect the
preceding literature fails to capture is \emph{user churn}. That is,
customers can switch service providers, if offered superior service
quality elsewhere. This aspect introduces \emph{competition} between the service providers (or coalitions of service providers) over market share, and turns the game into a \emph{partition form game} (described below). To the best of our knowledge, the interplay between resource pooling among service providers (aided by the associated economies of scale) and the competition between them, in the context of congestible services, has not been explored in the literature. This paper seeks to fill this gap.}{} 

This paper also contributes to the theory of coalition formation games 
in terms of new notions of stability. In particular, we focus on partition form games; the main ingredients of such games are, a partition (an arrangement of players into disjoint coalitions), the worth of each coalition (which, crucially, also depends on the partition), and the anticipation rules by which a  blocking or opposing coalition estimates its new worth (depending upon the anticipated retaliation of the opponents). 
In such games, the classical notion of stability declares a partition to be stable if it is not blocked by \textit{any} coalition (\cite{aumann1961,narahari})---a coalition blocks a partition if it anticipates 
greater worth in the new arrangement. However, some case studies may have no stable partitions under such classical notions (e.g., the game studied in~\cite{Shiksha_Perf}, and the market-size driven coalition formation game of the present paper). This necessitates a deeper study of such scenarios, possibly using new, more relevant notions of stability. \textit{In this paper, we define novel notions of stability by suitably restricting the set of candidate blocking coalitions.} Indeed, in practice, rearrangements in the marketplace typically arise from mergers between, or the breaking up of, existing corporations---our new notions of stability restrict the focus only on such tensions in the marketplace. 

\new{In this paper, we analyse a coalition formation game  between a collection of service providers, each of which is modelled as an Erlang-B loss system. A key aspect of our model is that the total market size (captured via the aggregate arrival rate of customer requests) is fixed exogenously, and providers (or coalitions of providers) compete for market share---this leads to a \emph{constant sum, partition form game}. These aspects, as we show, dramatically alter the outcome of the strategic interaction between providers.}{}
Interestingly, we find that under classical notions of stability\textit{, no arrangement of service providers into coalitions is stable, no matter how the payoff of each coalition is distributed across its members}. However, we demonstrate stable partitions when blocking coalitions are restricted to mergers and splits of the existing coalitions. Under our new notions of stability (we define two new notions, that differ on how a blocking coalition estimates its worth), \emph{the grand coalition is not} stable, except in a very specific corner case. \new{Instead, the predominantly stable configurations are \emph{duopolies}, with the larger coalition exploiting economies of scale to corner a disproportionate portion of the market size. Our work also highlights several subtleties relating to different natural notions of stability in this context, the way the payoff of each coalition is divided between its members, and the degree of congestion in the system.
}{}

\subsection*{Our contributions}

$\bullet$ \new{We formally define a constant sum coalition formation
game between strategic service providers of a congestible
service (see Section~\ref{sec_model}). This model is the first, to the best of our knowledge, to
capture the interplay between resource pooling and
competition over market share.}{} 


$\bullet$ 
Under the classical notion of stability for this \textit{partition form game} model (inspired by~\cite{aumann1961}), which we refer to as \textit{General Blocking-Perfect Assessment}, we show that no configuration is stable (see Section~\ref{sec:classical}). (A configuration specifies a partition of the set of providers into coalitions, and also the allocation of the total payoff of each coalition among its members.) This is because of the vast (specifically, all possible) range of deviations that can challenge any given configuration.

$\bullet$ In view of this impossibility result, \new{we define two novel \textit{restricted} notions of stability (see Section~\ref{stable_config}), where only coalitions arising from mergers or splits of existing coalitions can challenge the status quo. 
The two notions differ with respect to the precision with which the coalition that seeks to `break' from the prevailing configuration can estimate the benefit from doing so. 

Interestingly, we show that our restricted notions of stability do admit stable configurations. Moreover, these stable configurations involve \emph{duopolies}, i.e., two competing coalitions (except for a certain corner case where the grand coalition is also stable). Intuitively, configurations involving three or more coalitions are unstable because economies of scale incentivize mergers of two or more (but not all) coalitions. On the other hand, the constant sum nature of the game dis-incentivizes the formation of a grand coalition (except in the corner case mentioned above).}{}



$\bullet$ Finally, we explore the impact of the overall congestion level on the stable duopolies, by analysing light and heavy traffic regimes (see Section~\ref{sec_two_partition}). All duopolies are stable in heavy traffic, whereas only duopolies with nearly matched service capacities are stable in light traffic.


\subsection*{Related Literature}

This paper is related to two distinct strands of literature: (i) the literature on coalition formation for resource pooling in queueing networks, and (ii) the literature on partition form games.

\emph{Resource pooling in queueing networks:} This literature is quite vast, and we only provide a brief survey here; a comprehensive review can be found in~\cite{karsten}. One line of this literature models each coalition as a single server queue. The service rate of each coalition is either assumed to be optimized by the coalition itself (see, for example,~\cite{gonzalez,garcia,yu}), or simply taken to be the sum of the intrinsic service rates of the members (see, for example,~\cite{anily2010,timmer,anily2011,anily2014}.
Another line of literature treats each coalition as a multi-server loss system--\cite{karsten2012} considers the case where the number of servers with each player is fixed apriori, and \cite{ozen,karsten2014} consider the case where a coalition optimizes the number of servers it operates.
%
%
Finally,~\cite{karsten} analyses the setting where each coalition is an $M/M/s$ queue (Erlang~C); they consider both the above mentioned models for the service capacity of a coalition.


All the above mentioned papers assume that each service provider has a dedicated customer base (modeled via an exogenously determined arrival rate of service requests). From a game theoretic standpoint, this simplification ensures that the worth/utility of each coalition depends only the members of that coalition. In contrast, in the present paper, we explicitly model \emph{user churn}, which induces competition between coalitions, and turns the game into a \emph{partition form} game, wherein the worth/utility of a coalition also depends on the arrangement of players outside that coalition.


\emph{Partition form games:} The earliest work in this area can be found in~\cite{aumann1961}. The authors define a general definition of cooperative games which is applicable to both characteristic and partition form games (without using these names). The term ``partition form game" was first coined in~\cite{lucas}, where the authors further develop the theory of this class of games.~\cite{aumann1974cooperative} extends various existing stability notions for characteristic form games to partition form games.

Majority of the literature on cooperative games deals with the stability of the grand coalition in characteristic form games. In contrast, there is only a limited literature on partition form games.~\cite{hafalir} established the conditions under which the grand coalition is stable for convex partition form games. The authors in~\cite{saad_unilateral} (spectrum sensing and access), \cite{Shiksha_Perf} (Kelly's mechanism) show that certain finer partitions other than the grand coalition can be stable against unilateral deviations for partition form games, while the authors in~\cite{bloch,yi} show the same for the classical notions of stability against coalitional deviations. The authors in~\cite{Shiksha_Perf} also study stability against coalitional deviations to show that the  grand coalition is stable when players are significantly asymmetric, while  no partition is stable when the players are identical. Finally,~\cite{ray} considers a dynamic coalition formation game and shows that finer partitions can emerge at the sub-game perfect equilibrium.


\section{Model and Preliminaries}
\label{sec_model}

\new{In this section, we describe our system model for coalition formation
between strategic service providers, characterize the behavior of the
customer base in response to coalition formation between service
providers, and introduce some background. 

\subsection{System model}
Consider a system with a set~$\mathcal{N} = \{1,\cdots,n\}$ of
independent service providers (a.k.a., agents), with provider~$i$
having~$N_i$ servers. Without loss of generality, we assume~$N_i \ge N_{i+1}$ for~$1 \leq i \leq n-1.$ All servers are identical,
and assumed to have a unit speed, without loss of generality. The
providers serve a customer base that generates service requests as per
a Poisson process of rate~$\Lambda.$ Jobs sizes (a.k.a., service
requirements) are i.i.d., with~$J$ denoting a generic job size, and~$\mathbb{E}[J] = 1/\mu.$

Service providers are strategic, and can form
coalitions with other service providers to enhance their
rewards. Formally, such coalition formation between the service
providers induces a partition~$\P = \{C_1,C_2,\cdots,C_k\}$ of~$\mathcal{N},$ where~$\cup_{i = 1}^kC_i = \mathcal{N}, \ C_i \cap C_j = \emptyset \text{ for all } i
\neq j.$ We refer to such a partition with~$k$ coalitions as a~$k$-partition. (Naturally, the baseline scenario where each service
provider operates independently corresponds to an~$n$-partition.)

In response to a partition~$\P$ induced by coalition formation between
service providers, the arrival process of customer requests gets split
across the~$k$ coalitions in~$\P$, with the arrival process seen by
coalition~$C$ being a Poisson process of rate~$\lambda^{\P}_{C},$
where~$\sum_{C \in \P} \lambda^{\P}_{C} = \Lambda.$ (We characterize
the split~$(\lambda^{\P}_{C},\ C \in \P)$ as a Wardrop equilibrium;
details below.) Each coalition~$C$ operates as an~$M$/$M$/$N_{C}$/$N_{C}$ (Erlang-B) loss system, with~$N_{C} = \sum_{j
  \in C} N_j$ parallel servers, and arrival rate~$\lambda^{\P}_{C}.$
This means jobs arriving into coalition~$C$ that find a free server
upon arrival begin service immediately, while those that arrive when
all~$N_{C}$ servers are busy get dropped (lost). Given the well known
insensitivity property of the Erlang-B system, the steady state
blocking probability associated with coalition~$C$ (the long run
fraction of jobs arriving into coalition~$C$ that get dropped),
denoted~$B_{C}^{\P},$ is given by the Erlang-B formula:
 \begin{align}
\label{Eqn_PB}
B_{C}^{\P} = B(N_{C},a^{\P}_{C}), \text{ where } a^{\P}_{C} := \frac{\lambda^{\P}_{C}}{\mu} \text{ and } B(M,a) = \frac{ \frac{a^{M}}{M!}   }{ \sum_{j=0}^{M} \frac{a^{j}}{j!} }. 
\end{align}

\subsection{User behavior: Wardrop equilibrium}

Next, we define the behavior of the customer base in response to
coalition formation across service providers, via the split~$(\lambda^{\P}_{C},\ C \in \P)$ of the aggregate arrival process of
service requests across coalitions. This split is characterized as a
Wardrop equilibrium (or WE; see~\cite{WE}).

In the context of our model, we define the WE split of the arrival
process of service requests across coalitions, such that the steady
state blocking probability associated with each coalition is equal.
Note that since the blocking probability associated with an `unused'
coalition would be zero, it follows that all coalitions would see a
strictly positive arrival rate. Thus, the WE (if it exists) is
characterized by a vector of arrival rates~$(\lambda^{\P}_{C},\ C \in \P)$ satisfying
\begin{equation}
  \label{Eqn_WE_properties}
  B^{\P}_C = B\left(N_C,\frac{\lambda_C^\P}{\mu}\right) = B^*  \ \forall\ C \in \P \text{ and }
 \sum_{C \in  \P } \lambda_C^\P  = \Lambda ,
\end{equation}
where~$B^*$ is the common steady state blocking probability for each
coalition. For any given partition~$\P,$ the following theorem
establishes the existence and uniqueness of the WE, along with some
useful properties (proof in Appendix~\ref{appendix_B}). 
\begin{theorem}
  \label{Thm_WE}
Given any partition~$\P$ between the service providers and market size~$\Lambda$, there is a
  unique Wardrop equilibrium~$(\lambda^{\P}_{C},\ C \in \P),$ where~$\lambda^{\P}_{C} > 0$ for all~$C \in \P,$ that satisfies~\eqref{Eqn_WE_properties}. Additionally, the following properties hold:
  \begin{enumerate}[(i)]
      \item For each~$C \in \P, \lambda_{C}^\P$ is a strictly increasing
  function of the total arrival rate~$\Lambda.$
  \item  If the partition~$\P'$ is formed by merging two coalitions~$C_i$ and~$C_j$ in partition~$\P$ where~$C_i \cup C_j \neq \mathcal{N}$ (with all other coalitions in~$\P$
  remaining intact), then~$\lambda^{\P'}_{C_i \cup C_j} > \lambda^{\P}_{C_i} +
  \lambda^{\P}_{C_j}.$
  \item If~$\P = \{C_1,C_2\},$ with $N_{C_1} > N_{C_2},$
  then~$\frac{\lambda^{\P}_{C_1}}{N_{C_1}} > \frac{\Lambda}{N} >
  \frac{\lambda^{\P}_{C_2}}{N_{C_2}}, \text{ where } N = \sum_{i \in
    \mathcal{N}} N_i.$
  \end{enumerate} 
 \end{theorem}
 
 Aside from asserting the uniqueness and strict positivity of the
 Wardrop split, Theorem~\ref{Thm_WE} also states that equilibrium
 arrival rate of each coalition is an increasing function of the
 aggregate arrival rate~$\Lambda;$ see Statement~$(i).$ Additionally,
 Statement~$(ii)$ demonstrates the statistical economies of scale due
 to a merger between coalitions: the merged entity is able to attract
 an arrival rate that exceeds the sum of the arrival rates seen by the
 two coalitions pre-merger. Finally, Statement~$(iii)$ provides
 another illustration of statistical economies of scale for the
 special case of a 2-partition---the larger coalition enjoys a higher
 offered load per server than the smaller one.

 \subsection{Coalition formation game: Preliminaries}
 
 Having defined the behavior of the user base, we now provide some
 preliminary details on the coalition formation game between the
 service providers.

 Recall that each service provider is strategic, and only enters into
 a coalition if doing so is beneficial.
 Given a partition~$\P$ that describes the coalitions formed by
 the service providers, we define the value or payoff of each
 coalition~$C \in \P$ to be~$\beta \lambda_C^{\P},$ where~$\beta > 0.$
 This is natural when~$\lambda_C^{\P}$ is interpreted as being
 proportional to the number of subscribers of coalition~$C,$ with each
 subscriber paying a recurring subscription fee. Without loss of
 generality, we set~$\beta = 1.$
 
 The value~$\lambda_C^{\P}$ of each coalition~$C$ must further be
 apportioned between the members of the coalition. Denoting the payoff
 of agent~$i$ by~$\phi_i^{\P},$ we therefore have~$\sum_{i \in C}
 \phi_i^{\P} = \lambda_C^{\P} \text{ for all } C \in \P.$
 Since the providers are selfish, they are ultimately interested only
 in their individual payoffs.
 Thus, the coalition formation between providers is driven by the
 desire of each provider to maximize its payoff, given the statistical
 economies of scale obtained via coalition, and also the
 \emph{constant sum} nature of this game (the sum total of the payoffs
 of all providers equals~$\Lambda$). Thus, the relevant fundamental
 questions are:
\begin{enumerate}
\item Which partitions can emerge as a result of the strategic
  interaction between providers, i.e., which partitions are stable? 
  Indeed, a precursor to this question is: how does one define a
  natural notion of stability?
\item It is apparent that the answer to the above question hinges on
  how the value of each coalition is divided between its
  members. Thus, a more appropriate question is: which coalitional arrangement of agents and subsequent division of the coalitional shares results in  stable configurations?
\end{enumerate}
Our aim in this paper is to answer these questions; such problems can
be studied using tools from cooperative game theory. 
%
In the next section, we begin with classical notions of stability and `blocking by a  coalition', available in the literature; we will observe that there exists no partition which is stable under these classical notions. In the later sections, we refine the notion of stability (using some form of restricted blocking) and study the configurations that are stable.}

\section{Classical Notions of Coalitional Blocking and Stability }
\label{sec:classical}

It is well known that non-partition type transferable utility cooperative games are characterized by tuple~$(\mathcal{N},  \nu)$, where~$\nu(C)$ for any subset~$C \subset \mathcal{N}$ represents the utility of coalition~$C$.
However, this is not sufficient for a partition form game, where a coalition's utility depends not only on the coalition's players but also on the arrangement of other players.
In this case~$\nu(C)$ (more appropriately)  can be defined as the set of  payoff vectors (of dimension~$n$) that are  anticipated to be achievable by the players of the coalition~$C$ (e.g.,~\cite{aumann1961}); and this anticipation is based on their expectation of the reactions of the agents outside the coalition. 
The stability concepts (e.g., core) are extended to these type of games (e.g.,~\cite{aumann1961}), which are discussed at length in Appendix~\ref{appendix_A}. In this section we discuss the same ideas in our  notations, in particular, we consider the notion of $\alpha$-efficient  $R$-core defined in~\cite{aumann1961} (more details  are in Appendix~\ref{appendix_A}). 

This notion of stability  is interlaced with the notion of a partition
(more precisely, a configuration defined below) being \emph{blocked} by some
coalition. \new{We begin with relevant  definitions.
Given a partition~$\P = \{C_1,\cdots,C_k\},$ the set of payoff
vectors consistent with~$\P$ is defined as:
$${\bm \Phi}^{\P} := \left  \{\Phi = [\phi_1,\cdots, \phi_n] \in \mathbb{R}^n_+: \ \sum_{j \in C_i} \phi_j = \lambda^{\P}_{C_i}\ \forall \ i \right \}.$$
\textit{A \textbf{configuration} is defined as a tuple~$(\P,\Phi),$ such that~$\Phi \in {\bm \Phi}^{\P}.$} 
\newline
Note that a configuration specifies not just a
partition of the agents into coalitions, but also specifies an
allocation of payoffs within each coalition, that is consistent with
the partition.

\textit{\textbf{Blocking   by a coalition:} A configuration~$(\P,\Phi)$ is \emph{blocked} by a coalition~$C \notin \P$ if, for any
partition~$\P'$ containing~$C,$ there exists~$\Phi' \in {\bm \Phi}^{\P'}$
such that~$
  \phi_j' > \phi_j \text{ for all } j \in C.
$}

 Basically, a new coalition can block an existing configuration,
  if each one of its members can derive strictly better payoff from this
  realignment (irrespective of the responses of the opponents in~$C^c$).
Equivalently, $(\P,\Phi)$ is blocked by coalition~$C \notin \P$ if,
for any partition~$\P'$ containing~$C,$ we have
$\lambda^{\P'}_C > \sum_{j \in C} \phi_j.$
Note that the above equivalence hinges on the \emph{transferable
  utility assumption inherent in our cooperative game,  by virtue of  which (partial) utilities can be transferred across agents.} Intuitively, a
coalition~$C \subset \mathcal{N}$ blocks configuration~$(\P,\Phi)$, if
the members of~$C$ have an incentive to `break' one or more coalitions
of~$\P$ to come together and form a new coalition. In particular, it
is possible to allocate payoffs within the blocking coalition~$C$ such
that each member of~$C$ achieves a strictly greater payoff,
irrespective of any (potentially retaliatory) rearrangements among
agents outside~$C.$
This is referred to in the literature as a \textit{pessimistic
  anticipation rule} (see~\cite{pessimistic, Shiksha_Perf} and Appendix~\ref{appendix_A}) or $\alpha$-efficient rule in~\cite{aumann1961}.
 
We refer the above pessimal anticipation based blocking  as GB-PA (General Blocking--Perfect Assessment) rule,  we first provide the precise summary: 

\textit{\textbf{GB-PA rule:} Under this rule, a configuration~$(\P,\Phi)$ is
blocked by \emph{any} coalition~$Q \notin \P$  if
\begin{equation}
  \label{eq:blocking_PA}
  \ulam_Q > \sum_{i \in Q} \phi_i, \text{ where }
  \ulam_Q := \min_{\P': Q \in \P'} \lambda^{\P'}_Q.
\end{equation}A configuration is stable under the GB-PA rule if it is not blocked by any coalition.}

The term `General Blocking' is used for this notion, as any arbitrary coalition (mergers or splits of the existing coalitions or mergers of partial splits) can block; and the term `Perfect Assessment' is used as the players in blocking coalition are aware of the previous shares of all members of the blocking coalition, i.e., previous shares of players is `common knowledge' within~$Q$.}{}

\textbf{Stability under GB-PA:} We establish a negative result for this classical notion of stability (proof in Appendix~\ref{appendix_C}):
\begin{theorem}
 For $n>2$, there exists no stable configuration under GB-PA rule.
\label{thm_impossible}
\end{theorem} 

 We establish the above result by showing that the configuration with the~$n$-partition (i.e., each agent operates alone) is blocked by a suitable merger, while for any other configuration, there exists a~$j \in \mathcal{N}$ such that either~$\{j\}$ or~$\mathcal{N}-\{j\}$ blocks it. For $n=2$ it is trivial to observe that the only stable configurations are $(\P_2,\Phi^{\P})$  and $(\{1,2\},\Phi^{\P})$ where $\P_2:=\{\{1\},\{2\}\}$ and $\Phi^{\P}:= (\lambda_{\{1\}}^{\P_2},\lambda_{\{2\}}^{\P_2})$.

Theorem~\ref{thm_impossible} states that no  configuration is stable under GB-PA for~$n>2$, in other words, the $\alpha$-core (R-core under $\alpha$-effectiveness) as defined in~\cite{aumann1961} is empty, for our game. This `impossibility' is due to the fact that under GB-PA, a configuration can be blocked by \emph{any} coalition that is not contained in it; this coalition can be formed via multiple mergers/splits of existing coalitions.  But in practice, either an existing coalition splits or two or more of the existing coalitions merge. Thus, to define more practical and relevant notions of stability, one may have to consider a more restricted set of blocking candidates. This is addressed in the next section.
 
 In the next section,  
we
also consider an alternate form of restricted blocking, where the
`prevailing worth' of the agents of the candidate blocking coalition
is assessed imprecisely. Prior to that, we conclude this section with a short discussion on other anticipation rules.

\textbf{Other Anticipation Rules:} There are many other anticipation rules considered in the literature, for e.g.,~$\beta$-effective rule in~\cite{aumann1961} (coalition~$C$ can block payoff vector~$\Phi$, if for every correlated strategy of players in~$\mathcal{N}-C$, there exists a correlated strategy of players in~$C$ which leaves them better-off) and max rule in~\cite{pessimistic} (the  opponents/players in~$\mathcal{N}-C$ are anticipated to arrange themselves in a partition that maximizes their own utilities).  
  Interestingly, the pessimistic rule coincides with the above mentioned anticipation rules for our constant sum game, mainly because of economies of scale established in Theorem~\ref{Thm_WE}.$(ii)$.
  
  There are other anticipation rules that do not coincide with the pessimal rule. For example, the optimistic rule (opponents are anticipated to arrange in such a way that the deviating coalition obtains the best utility) in~\cite{pessimistic}, the Cournot Nash Equilibrium (opponents are anticipated to remain in their old coalitions) in~\cite{alpha-core}, etc. However, the impossibility result established in Theorem~\ref{thm_impossible} also implies impossibility under these rules (if any coalition $Q$ anticipates a higher utility than what its members derive in the current configuration under the pessimal rule~\eqref{eq:blocking_PA}, it  would also anticipate higher utility using any other anticipation~rule). 
  

\section{Realistic Notions of Blocking and Stability} 
\label{stable_config}

Motivated by the impossibility of stable configurations under GB-PA (Theorem~\ref{thm_impossible}), in this section, we define weaker, more realistic notions of stability, that do admit stable configurations. Specifically, the proposed stability notions
differ from GB-PA on the class of candidate blocking
coalitions considered, as well as the precision with which the
`prevailing worth' of the members of the candidate coalition is
assessed and/or revealed. The former distinction is inspired by the observation that organisational rearrangements predominantly occur in practice via mergers or splits of existing coalitions. For each of these notions of stability, we characterize the
class of stable configurations. 

\new{The main takeaway from our results is the following.
 Because of the interplay between statistical economies of scale and the
constant sum nature of the game, only configurations involving duopolies (i.e., partitions with two coalitions) are stable (except in a certain corner case, where the grand coalition is also stable). This is true for both the proposed notions of stability defined next. }{}

\subsection{Restricted blocking and stability}

\new{The first notion of stability we introduce simply restricts the set of
candidate blocking configurations to mergers and splits of prevailing
coalitions. Note that this is a natural restriction from a practical
standpoint, since complex rearrangements between firms in a
marketplace typically arise (over time) from a sequence of mergers and
splits. We refer to this as restricted blocking (RB). Further when one
assumes the precise knowledge of the worth of the blocking candidates,
it leads to the RB-PA (Restricted Blocking--Perfect Assessment)
rule. We begin with this rule.

\textit{\textbf{RB-PA rule:} Under this rule, a configuration~$(\P,\Phi)$ can be blocked only by a coalition~$Q$ that is formed either via~$(i)$ a merger of
coalitions in~$\P$ (i.e.,~$Q = \cup_{C \in \mathcal{M}} C$ for~$\mathcal{M} \subseteq \P$), or via ~$(ii)$ a split of a single coalition in~$\P$ (i.e.,~$Q \subset C$ for some~$C \in \P$). Further, such a~$Q$ blocks~$(\P,\Phi)$ if, for all
partitions~$\P'$ containing $Q,$ there exists
~$\Phi' \in {\bm \Phi}^{\P'}$ such that~$ \phi'_i > \phi_i \text{ for all } i \in Q.$}

\textit{Equivalently,~$Q$ (as described above) blocks the configuration~$(\P,\Phi)$ if
\begin{equation}
  \label{eq_def_RBPA}
  \ulam_Q > \sum_{i \in Q} \phi_i, \text{ where }
  \ulam_Q := \min_{\P': Q \in \P'} \lambda^{\P'}_Q.
\end{equation}
A configuration~$(\P,\Phi)$ is stable under the RB-PA rule if it is
not blocked by any merger or split.}

Note that like GB-PA, the RB-PA rule also involves pessimal anticipation; the
members of candidate blocking coalition are pessimistic in their
anticipation of the value of the new coalition. Moreover, it is
possible to allocate the payoff of the blocking coalition~$Q$ among its members such that each member is (strictly) better off, as discussed in the previous section. 

The next  notion  uses the same restriction on the set of
candidate blocking coalitions, but uses an imprecise assessment of
the prevailing worth of the members of the candidate blocking
coalition, resulting in an imprecise  assessment of the benefit/loss from blocking. We refer to this as the
RB-IA (Restricted Blocking--Imperfect Assessment) rule.}{}

\textit{\textbf{ RB-IA rule:}
Under this rule, a configuration~$(\P,\Phi)$ is blocked by a coalition~$Q$ formed either via a merger or a split if:
\begin{align}
  \label{Eqn_unified_condition_S}
  &\ulam_Q := \min_{\P': Q \in \P'} \lambda^{\P'}_Q > \sum_{C \in \P} \frac{N_{C \cap Q}}{N_C} \lambda^{\P}_C, \\
  \label{Eqn_unified_condition_S_pt2}
  &\lambda^{\hat{\P}}_Q > \sum_{i \in Q} \phi_i, \text{ where }\hat{\P} =  \left( \bigcup_{C \in \P} \{ C\setminus Q\} \right) \bigcup \{Q\}.
 \end{align}
 A configuration~$(\P,\Phi)$ is stable under the RB-IA rule if it is not blocked by any merger or split.}
 
\new{ Condition~\eqref{Eqn_unified_condition_S} can be
interpreted as a first stage check on the feasibility of the block, by
(imperfectly) assessing the total prevailing worth of the members
of~$Q$ (using the prevailing coalitional worths~$\{\lambda_C^\P\}$). This imprecise assessment is obtained as the sum of the proportional contributions of the members of~$Q$ to their respective parent coalitions; the imprecision stems from not using the actual payoffs~$\{\phi_i\}_{i \in Q}$.  
Note that this feasibility check is also under the pessimal anticipation rule, but with imperfect estimates. 

Condition~\eqref{Eqn_unified_condition_S_pt2} is the final validation of the block using precise estimates~$\{\phi_i\}_{i \in Q}$. This ensures that it is possible to
allocate the payoff of~$Q$ among its members such that each member is
(strictly) better off from the deviation. Here, the anticipation is that there would be no immediate retaliation from the leftover players, i.e., as seen from the definition of~$\hat{\P}$ in~\eqref{Eqn_unified_condition_S_pt2}, the opponents would remain in their
original coalitions (as in the Cournot Nash equilibrium~\cite{alpha-core}).}{} This is reasonable after the already pessimal feasibility check in~\eqref{Eqn_unified_condition_S}.

Let us now interpret the condition for blocking due to a split/merger separately under RB-IA. We begin with blocking due to a split.
\new{By~\eqref{Eqn_unified_condition_S} and~\eqref{Eqn_unified_condition_S_pt2}, a configuration~$(\P,\Phi)$ is
blocked by a coalition~$Q$ that is formed by splitting a coalition~$C \in \P$ if: \vspace{-2mm}
\begin{align}
  \label{Eqn_condition_S}
  &\ulam_Q := \min_{\P': Q \in \P'} \lambda^{\P'}_Q > \frac{N_Q}{N_C} \lambda^{\P}_C, \\
  \label{Eqn_condition_S_pt2}
  &\lambda^{\hat{\P}}_Q > \sum_{i \in Q} \phi_i, \text{ where }\hat{\P} = (\P \setminus \{C\}) \cup \{Q, C\setminus Q\}.
 \end{align}
Condition~\eqref{Eqn_condition_S} estimates the total prevailing worth of the members
of~$Q$, as proportional to their fractional contribution towards the service capacity
of~$C$, i.e.,~$N_Q/N_C$. Condition~\eqref{Eqn_condition_S_pt2} is the final
stage check on split feasibility as discussed above. Note that~$\hat{\P}$ is the new partition that emerges after the split when opponents remain in their original coalitions. 

Applying~\eqref{Eqn_unified_condition_S} and~\eqref{Eqn_unified_condition_S_pt2} to a merger, 
a configuration~$(\P,\Phi)$ is
blocked by a merger coalition~$Q = \cup_{C \in \mathcal{M}} C, \text{ for some } \mathcal{M} \subseteq \P$, 
if 
\begin{equation}
  \label{Eqn_condition_M}
\ulam_Q  >  \sum_{C \in \mathcal{M}} \lambda^\P_C \mbox{, and } \lambda^{\hat{\P}}_Q > \sum_{i \in Q} \phi_i, \text{ where } \hat{\P} = \{Q, \P \backslash \mathcal{M} \}.
\end{equation}
Note that the first condition in~\eqref{Eqn_condition_M} is identical to~\eqref{Eqn_unified_condition_S}, the only difference being that the prevailing worth of all the deviating members~$(\sum_{C \in \mathcal{M}}\lambda_C^\P)$ is assessed precisely, given that full coalitions are deviating. The second condition in~\eqref{Eqn_condition_M} is the same as~\eqref{Eqn_unified_condition_S_pt2}.
However, observe   that~$\sum_{i \in Q} \phi_i =  \sum_{C \in \mathcal{M}} \lambda^\P_C,$ and hence the second condition in~\eqref{Eqn_condition_M} is implied by the first, as~$\ulam_Q  \le \lambda^{\hat{\P}}_Q$.

Note that RB-PA and RB-IA differ only in the condition for blocking
due to a split. This is natural, since the net worth of coalitions
 $\{\lambda_C^\P\}_{C \in \P}$ is often common knowledge,
whereas the internal payoff allocation within a coalition can often be
confidential.

Having defined our new notions of stability, we now consider each notion
separately, and characterize the resulting stable configurations. We
begin with RB-IA, which appears to admit a broader class of stable configurations.}{}

\subsection{Stable configurations under RB-IA}

\new{Our first result is that all configurations involving partitions
of size three or more are unstable. In other words, only monopolies or
duopolies can be stable (proof in Appendix~\ref{appendix_C}).

\begin{theorem}
\label{Thm_duo_mono}
\textit{Under the RB-IA rule, any configuration~$(\P,\Phi)$ with~$|\P| \geq 3$ is not stable. } 
\end{theorem}}{}

The proof sheds light on why
configurations with~$|\P| \geq 3$ are unstable -- they are blocked by any merger leading to a~$2$-partition; this is because of the economies of scale arising from such a merger (as shown in Theorem~\ref{Thm_WE}.$(ii)$), and the pessimal anticipation rule.

\new{Next, we move to the two remaining possibilities:
stable configurations involving the grand coalition, and those
involving 2-partitions.

 {\bf Grand Coalition:} Defining~$\P_G := {\cal N}$ as the
grand coalition, it is clear that any configuration of the
form~$(\P_G,\Phi)$ can only be blocked by a split. 
 We now show that
unless a single agent owns at least half the total service capacity
of the system, such a block is always possible. In other words, any
configuration involving the grand coalition is unstable, unless there
is a single `dominant' agent. On the other hand, if there is a single
agent who owns at least half the service capacity, we show that there
exist stable configurations of the form~$(\P_G,\Phi)$
(see Appendix~\ref{appendix_C} for proof).

\begin{theorem}
  \label{Thm_GC} 
Under the RB-IA rule: 
\begin{enumerate}[(i)]
    \item If~$N_1 < \sum_{ i \neq 1}N_i$, then
  there exists no payoff vector $\Phi$ consistent with ${\P}_G$, such
  that $({\P}_G, \Phi)$ is stable.
  \item If $N_1 \ge \sum_{ i \neq 1}N_i$, then there
  exists at least one payoff vector $\Phi$ consistent with ${\P}_G$, such
  that $({\P}_G, \Phi)$ is stable. Specifically, any configuration $({\P}_G, \Phi)$  satisfying the following is stable:  
\begin{equation}
   \phi_1 \ge \max \left  \{ \ulam_C: C \subsetneq {\cal N} \mbox{ and } 1 \in C \right \}. 
   \label{Eqn_payoff_RB_GC}
\end{equation}
\end{enumerate} 
\end{theorem}}{}
To prove part~$(i)$ of the above theorem, we show that for any payoff vector, there exists a coalition with~$n-1$ players that blocks the grand coalition (details in Appendix~\ref{appendix_C}).
For part~$(ii)$, note that only coalitions containing player~$1$  satisfy condition~\eqref{Eqn_condition_S} and hence are  potential blocking coalitions under RB-IA. Therefore, if player~$1$ is given a large enough allocation (as in~\eqref{Eqn_payoff_RB_GC}) in the grand coalition, it does not have an incentive to deviate, either alone or as part of a group.

\new{{\bf Duopolies:} We are now left to examine the stability of duopolies, i.e.,~$2$-partitions, under the RB-IA rule.
Duopolies can, without loss of
generality, be represented as~$\P = \{C_1,C_2\},$ with~$k := N_{C_1} \geq
N_{C_2}.$ In the following, we identify a family of stable duopolies under RB-IA rule. An interesting property of the stable configurations we identify is that, the stability does not depend upon the payoff vector,~$\Phi$. Instead, it only depends upon the specifics of the partition (however this is not true for all partitions).  
\textit{This insensitivity 
to the payoff vector is not seen under the RB-PA rule.}
We begin by defining some preliminaries.

\textit{\textbf{Stable partition:}
  A partition~$\P$ is stable if all configurations
  involving it are stable, i.e., configuration~$(\P, \Phi)$ is stable
  for any~$\Phi \in {\bm \Phi}^\P.$}

 By
Theorem~\ref{Thm_WE},~$\lambda_{C_1}^\P = \lambda^\P_k$ is the unique zero of the following
function (see~\eqref{Eqn_WE_properties}):
\begin{equation}h(\lambda) := \frac  { \lambda^k } { k! } \sum_{j=0}^{N-k} \frac{
  (\Lambda-\lambda)^j }{j!}  -
\frac{{(\Lambda-\lambda)}^{N-k}}{(N-k)!}  \sum_{j=0}^{k}
\frac{\lambda^j} {j!}. \nonumber\end{equation} 
Now, we define~$\Psi(k; \Lambda):= \lambda_k^\P/k$ as the offered load (or market size) per
server of the larger coalition. Finally, define
\begin{eqnarray}
\label{def_k_star}
  k^{*}(\Lambda) := {\arg \max}_{k}
  \Psi(k ; \Lambda).  \label{Eqn_kstar}
\end{eqnarray}
Note that~$k^*(\Lambda)$ is the set of values of~$k$ that maximizes
the per-server offered load of the larger coalition among~$2$-partitions.

Let~${\mathbb C}^* := \{ C \subsetneq {\cal N}: N_C \in k^* (\Lambda) \}
$ be the set of coalitions~$C$, that can derive the maximum per-server
offered load among~$2$-partitions. In the following lemma, we provide a sufficient condition for
a class of partitions (recall any such partition is represented by $\P = \{C_1,C_2\}$) to be stable. 

\begin{lemma}
\label{Lemma_stable_two_RB-IA}
 Consider the RB-IA rule. A~$2$-partition~$\P$ is stable
  if there exists no coalition~$S \subset C_i \text{ for }
  i=\{1,2\}$ such that
 $  \frac{ \ulam_S } {N_S } > \frac{\ulam_{C_i}}{N_{C_i}}= \frac{\lambda_{C_i}^{\P}}{N_{C_i}}. $
\end{lemma}

The proof of the lemma follows directly from the definition of stability. Indeed, for~$2$-partitions that satisfy the hypothesis of the above lemma, none of the splits  are feasible (they violate~\eqref{Eqn_condition_S}); further, the merger of both coalitions (which leads to grand coalition) is also not feasible because of the constant sum nature of the game. A consequence of this lemma is the following (see Appendix~\ref{appendix_C} for the proof).}{}

\begin{theorem}
  \label{Thm_two_partition}
\new{Consider the RB-IA rule.}{}
\begin{enumerate}[(i)]
    \item There always exists a stable~$2$-partition.
  \item \new{Any~$2$-partition~$\P$ with one
  of the coalitions from~$\mathbb {C}^*$ is a stable
  partition. }{}
  \item Additionally, any~$2$-partition~$\P= \{C_1,C_2\}$ (where~$N_{C_1} \ge N_{C_2}$) with no~$C \subsetneq C_1$ such that~$N_{C} > N/2$ is stable. 
    \end{enumerate}
\label{Thm_Prule_stability}
\end{theorem} 
Note that statement~$(i)$ directly follows from statement~$(ii)$ and the non-emptiness of $\mathbb{C}^*$. Statement~$(ii)$ follows as the duopolies identified here satisfy the hypothesis of Lemma~\ref{Lemma_stable_two_RB-IA}. A similar reasoning applies for statement~$(iii)$. 

From Theorem~\ref{Thm_Prule_stability}.$(iii)$, the duopolies with perfectly matched service capacities~($N_{C_1} = N_{C_2}$) are also stable; while from~$(ii)$ any duopoly with $N_{C_1} = k^*$ (see~\eqref{Eqn_kstar}) is stable.
\new{Further, Theorem~\ref{Thm_two_partition} identifies a class of
\emph{stable partitions}, i.e., partitions that are stable for any consistent payoff vector. However, there can also exist duopolies that are stable only under certain consistent payoff vectors and unstable for others}{} (see Section~\ref{sec_numerical}). 

In Section~\ref{sec_two_partition}, we provide a complete
  characterization of the class of stable partitions under RB-IA, in
 the heavy and light traffic regimes.

 \subsection{Stable configurations under RB-PA}
\label{sec_PA}

\new{Next, we consider stable configurations under the RB-PA rule. Under this rule, 
we show that only configurations involving~$2$-partitions can be stable,
i.e., configurations involving the grand coalition, or
involving~$k$-partitions with~$k \geq 3$ are always
unstable. In contrast, for the RB-IA rule, recall that the grand coalition  is stable   under certain conditions. Moreover, also in contrast to RB-IA, the stability/instability of duopoly configurations under RB-PA appears to depend on the associated payoff vector.}{}

We begin by characterising the space of stable allocations under RB-PA.
From \eqref{eq_def_RBPA},
it is easy to see that a stable payoff vector lies in the polyhedron~\eqref{eqn_polyhedra} defined below.
\begin{lemma}
  {\bf[Polyhedral Characterisation]} Given any partition~$\P$, stable allocations lie in the polyhedron defined by
  \begin{equation}
  \label{eqn_polyhedra}
      \sum_{i \in Q} \phi_i \geq \ulam_Q \mbox{ for all } Q \subseteq C_j \in \P \text{, and for all } j.
  \end{equation}
\end{lemma}

It is clear from the above lemma that RB-PA does not admit stable partitions (unlike RB-IA). In other words, stability under RB-PA is tied to the payoff vector. Interestingly, stable partitions under RB-IA, paired with a special payoff vector (defined next) are stable; see Theorem~\ref{Thm_stable_config_ruleB}.

\new{The \emph{proportional} payoff
vector $\Phi^{\P}_p,$
associated with any partition~$\P$, assigns to each member a payoff in proportion to the number of servers they bring to the
coalition: 
\begin{eqnarray}
\label{Eqn_PSA}
\phi^{\P}_{p, i}  =  \frac{N_i}{ \sum_{j \in C} N_j }   \lambda_C^\P \mbox{ for any } i  \in C \in \P.
\end{eqnarray}

Our results for the RB-PA rule are summarized as follows (see Appendix~\ref{appendix_C} for the proof).}{}
\begin{theorem}
  \label{Thm_stable_config_ruleB}
\new{Under the RB-PA rule:}{}
\begin{enumerate}[(i)]
    \item \new{No configuration involving the grand coalition is
  stable.}{}
  \item \new{No configurations involving~$k$-partitions, for~$k \geq 3$ are stable.}{}
  \item There exists at least one~$2$-partition~$\P$ such that~$(\P,\Phi_p^\P)$ is stable. \new{Specifically, consider any stable~$2$-partition~$\P$ under the RB-IA rule. Then~$(\P,\Phi_p^\P)$ is stable under the RB-PA rule. 
Further, there exists a neighbourhood~${\cal B}^\P_p$ of the payoff vector~$\Phi_p^\P$ such that~$(\P,\Phi)$ is stable for all~$\Phi \in {\cal B}^\P_p$.}{}
\end{enumerate}
 \label{Thm_R_rule_stability}
\end{theorem}
Like RB-IA, RB-PA also does not admit any stable configurations involving~$3$ or more coalitions. Moreover, under RB-PA, the grand coalition is  also unstable for all payoff vectors (unlike RB-IA, which admits payoff vectors that stabilise the grand coalition under certain conditions).
\new{Finally, turning to duopolies, Theorem~\ref{Thm_stable_config_ruleB} conveys that partitions that are
stable under the RB-IA rule (irrespective of the associated payoff
vector), are also part of stable configurations under RB-PA, but under
a restricted class of payoff vectors. Specifically, the payoff vectors
we identify are `close' to proportional allocations.}{}  
Next we investigate  
other natural payoff structures  that also induce stability under RB-PA. In particular, we consider a payoff vector inspired by the classical Shapley value.  

\textbf{Shapley value: }Shapley value  is one of the well-known sharing concepts used in cooperative game theory~(\cite{narahari}). 
We begin by defining an extended version of Shapley value for partition form games, to divide a coalition's worth among its members~(\cite{aumann1974cooperative}). Under this extension, we treat each coalition~$C_i$ in the partition as a `grand coalition', define a suitable `worth'~$\nu_C$ for each~$C \subset C_i$, and then use the usual definition of Shapley value to obtain individual shares of the players in~$C_i$. Formally, for any~$j \in C_i$ , 
\begin{eqnarray}
		\label{Eqn_SV}
		\phi_{s,j}^\P := \sum_{C \subseteq C_i, j \notin C} \frac{|C|!(|C_i|-|C|-1)!}{|C_i|!} \left[\nu_{C \cup \{j\}} - \nu_C \right], 
	\end{eqnarray} where $\nu_C$ is defined using \textit{pessimal anticipation} as below: 
 \begin{equation}
    \nu_C = \lambda_C^{\P'},  \text{ where } \P' = \P \backslash \{C_i\} \cup \{C, C_i \backslash C\}.   \label{eq_subcoalition_worth}
\end{equation}
Note that~$\nu_C$ is defined as the payoff obtained by~$C$ when (i) players outside of~$C_i$ remain attached to their original coalitions (as in the Cournot equilibrium), and (ii) the players in~$C_i \setminus C$ form a single competing coalition (in the spirit of pessimal anticipation).

Next, we present some contrasting results (compared to Theorem~\ref{Thm_R_rule_stability}) for a small number of service providers, for any~$2$-partition~$\P = \{C_1,C_2\}$ (proof in Appendix~\ref{appendix_C}).
  
\begin{theorem}
   \label{Thm_SV_small}
 Under the RB-PA rule, with the Shapley payoff vector~$\Phi^\P_s$ as defined in~\eqref{Eqn_SV} and~\eqref{eq_subcoalition_worth},
  \begin{enumerate}[(i)]
      \item for~$n=3$, the configuration~$(\P,\Phi_s^\P)$ is stable for any~$2$-partition~$\P$, and
      \item for~$n=4$, the configuration~$(\P,\Phi_s^\P)$ is stable for any~$2$-partition~$\P$ such that~$|C_1| = |C_2| = 2$. 
  \end{enumerate}
  \end{theorem} 
 Note that Theorem~\ref{Thm_SV_small} establishes the stability of certain~$2$-partitions under the Shapley payoff vector that are not covered in Theorem~\ref{Thm_R_rule_stability} under the proportional payoff vector (for~$n = 3,4$). Specifically, under the Shapley payoff vector, any~$2$-partition for~$n=3,$ and any~$2$-partition with equal-sized coalitions for~$n=4,$ is stable. In contrast, recall that the~$2$-partitions that are shown to be stable under the proportional payoff vector depend on the number of servers within each coalition (see Theorem~\ref{Thm_R_rule_stability}). We present a few examples in Section~\ref{sec_numerical} to demonstrate these contrasts numerically.

 \section{Stable Duopolies: Heavy and Light Traffic }
\label{sec_two_partition}

In this section, we provide a complete characterization of stable partitions under RB-IA, and stable configurations under RB-PA with the proportional payoff vector, in heavy and light traffic regimes. Specifically, we provide the necessary and sufficient conditions for stability, as~$\Lambda \uparrow \infty$ (heavy traffic) and~$\Lambda \downarrow 0$ (light traffic), with other system parameters remaining unchanged.
 
 Our analysis presents interesting contrasts between the heavy and light traffic regimes. In heavy traffic, we find that \textit{all} duopolies form stable partitions under RB-IA and stable configurations (with the proportional payoff vector) under RB-PA. Intuitively, this is because economies of scale discourage splits in heavy traffic; as we show in Lemma~\ref{lem_psi_inc} in Appendix~\ref{appendix_C}, the per server utility of the larger coalition \textit{increases} with the number of servers it possesses (Interestingly, this is a `second order' effect; per server scales as~$\nicefrac{\Lambda}{N}$ for both coalitions in heavy traffic (see Lemma~\ref{lem_accuracy_heavy_tight} in Appendix~\ref{appendix_C}).). In contrast, in light traffic, we find that only duopolies where the two coalitions are `closely matched' in the number of servers they possess, are found to be stable. Intuitively, this is because economies of scale get significantly diluted in light traffic, discouraging any coalition from becoming `too large'.

\subsection{Heavy Traffic}  
Our main result in heavy traffic is the following (proof in Appendix~\ref{appendix_C}).
\begin{theorem} 
\label{Thm_heavy}
There exists  a ~${\bar \Lambda}$ such that for all~$\Lambda \ge {\bar \Lambda}$, the following holds: given any~$2$-partition~$\P,$
 \begin{enumerate}[(i)]
     \item $\P$ is a stable partition under RB-IA, and
     \item $(\P,\Phi_p^\P)$  is a stable configuration under RB-PA. 
 \end{enumerate} 
 \end{theorem}
 This result can be interpreted as follows. Note that due to the constant sum nature of the game, duopolies can never be blocked due to a merger. Thus, our stability analysis hinges on the feasibility of splits. Specifically, we prove Theorem~\ref{Thm_heavy} by showing that given any~$2$-partition~$\P$,
 \begin{enumerate}[(a)]
     \item for any consistent payoff vector~$\Phi,$ the configuration~$(\P,\Phi)$ cannot be blocked by a split under RB-IA, and
     \item the configuration~$(\P,\Phi_p^\P)$ cannot be blocked by a split under RB-PA.
 \end{enumerate}
 These statements in turn follow from the fact that in heavy traffic, the per-server offered load~$\Psi(k)$ of the larger coalition increases monotonically with the number of servers~$k$ it possesses, i.e., its service capacity (see Lemma~\ref{lem_psi_inc} in Appendix~\ref{appendix_C}). In other words, economies of scale persist in heavy traffic. Indeed, the above monotonicity property, which is proved by exploiting the analytical extension of the Erlang-B formula to real-valued service capacities (see~\cite{jagerman}), renders condition~\eqref{Eqn_condition_S} for a split under RB-IA, and condition~\eqref{eq_def_RBPA} for a split under RB-PA, invalid.
	
\subsection{Light Traffic}
Next, we consider the light traffic regime and our main result here is  (proof in Appendix~\ref{appendix_C}):
\begin{theorem}
 Let~${\mathscr P}$ denote the space of~$2$-partitions~$\P = \{C_1,C_2\}$ (where~$N_{C_1} \ge N_{C_2}$) satisfying the following condition:
 there does not exist~$C \subset C_1$ such that~$N_{C} > N/2.$ There exists~${\underline \Lambda} >0$, such that for all~$\Lambda \le {\underline \Lambda}$,
  \begin{enumerate}[(i)]
  \item $\P$ is a stable partition under RB-IA  if and only if~$\P \in \mathscr{P}$, and
  \item $(\P,\Phi_p^\P)$ is a stable configuration under RB-PA  if and only if~$\P \in \mathscr{P}$.
  \end{enumerate}
   \label{low_traffic}
     \end{theorem}
Theorem~\ref{low_traffic} highlights that the~$2$-partitions that are stable under RB-IA and form stable configurations (with the proportional payoff vector) under RB-PA are those where the service capacities of the two coalitions are nearly matched. Formally, the larger coalition~$C_1$ should not have a sub-coalition~$C$ with more than half the total service capacity. In particular, note that duopolies with perfectly matched service capacities~($N_{C_1} = N_{C_2}$) also satisfy this condition. Intuitively, the result holds because in light traffic, the larger coalition corners almost the entire offered load (i.e.,~$\nicefrac{\lambda_{C_1}^\P}{\Lambda} \to 1 \text{ as } \Lambda \to 0$); see Lemma~\ref{lem_light_new} in Appendix~\ref{appendix_C}. 

Our results in the heavy and light traffic regimes shed light on the impact of congestion (via the total offered load~$\Lambda$, a.k.a., the \textit{market size}) on coalition formation. In light traffic, the per-server utility of the larger (by service capacity) coalition~$\Psi(k)$ \textit{decreases} with its service capacity~$k$ (as the larger coalition captures almost the entire~$\Lambda$, irrespective of~$k$). This in turn encourages duopolies where the service capacities of the two coalitions are closely matched (even though the larger coalition corners most of the total utility). On the other hand, in heavy traffic, the per-server utility of the larger (by service capacity) coalition~$\Psi(k)$ \textit{increases} with its service capacity~$k.$ These economies of scale induce stability in all duopolies, including those that have coalitions with highly asymmetric  service capacities. This suggests that in general, at moderate congestion levels, the per-server utility of the larger (as before, by service capacity) coalition peaks at an intermediate value of~$k$ between~$\nicefrac{N}{2}$ and $N,$ encouraging the formation of moderately asymmetric duopolies. This is consistent with what we find in our numerical experiments  (see Figure~\ref{fig:stable partitions}).

Finally, it is important to note that we are able to provide necessary and sufficient conditions for stability under RB-IA and RB-PA in heavy and light traffic regimes; in contrast, we could only   provide sufficient conditions for stability (see Theorems~\ref{Thm_two_partition} and~\ref{Thm_stable_config_ruleB}) outside of these limiting regimes.

\section{Numerical Case Studies}
\label{sec_numerical}
		
In this section, we present some numerical case studies that illustrate our key findings. 
Importantly, we also consider examples for which the conditions of our theorems are not satisfied; these provide additional insights. 
We numerically compute~${\lambda}_C^\P$  for various~$C$ and~$\P$ using zero finding algorithms and then compute~$k^*$ of~\eqref{Eqn_kstar} or  use equations~\eqref{Eqn_condition_S}-\eqref{Eqn_condition_S_pt2} or~\eqref{eq_def_RBPA} to determine the stable configurations.

 \textbf{RB-IA rule:} Recall that Theorem~\ref{Thm_two_partition} provides sufficient condition for a \textit{stable partition} under RB-IA, i.e., a partition that is stable under any consistent payoff vector. Here, we illustrate that RB-IA also admits stable configurations that are not supported by stable partitions. Consider the example with~$\Lambda = 13$ and~$4$ service providers having service capacities:~$N_1 = 10,$ $N_2 = N_3 = N_4= 2$. Note that the partition~$\P = \{\{1,2,3\},\{4\}\}$ does not satisfy the hypothesis of Theorem~\ref{Thm_two_partition} (in this case,~$k^* =\{12\})$. Moreover, configuration~$(\P,\Phi_p^{\P})$ is blocked by~$Q = \{1,2\}$ as split~$Q$ satisfies~\eqref{Eqn_condition_S}, while,~$\Phi_p^\P$ and~$Q$ satisfy~\eqref{Eqn_condition_S_pt2}. Thus, $\P$ is not a stable partition. However, the configuration $(\P,\Phi)$ is stable for the following set of payoff vectors:
\begin{eqnarray}
  \left \{\Phi : \phi_1 \ge \lambda_{\{1\}}^{\{\{1\},\{2,3\},\{4\}\}}, \ \phi_1+\phi_2 \ge \lambda_{\{1,2\}}^{\{\{1,2\},\{3\},\{4\}\}}, \  \phi_1 + \phi_3 \ge \lambda_{\{1,3\}}^{\{\{1,3\},\{2\},\{4\}}, \right. & \nonumber \\ 
 \left. \phi_2 + \phi_3 \ge  \lambda_{\{2,3\}}^{\{\{1\},\{2,3\},\{4\}\}}, \ \phi_1+\phi_2+\phi_3 = \ulam_{\{1,2,3\}}   \right \}. \nonumber 
\end{eqnarray}
It can be checked that this set is indeed non-empty.
This demonstrates that it is possible for a partition to be stable under some but not all consistent payoff vectors.
 
\begin{figure}[ht]
\begin{minipage}{.55\textwidth}
\vspace{-10mm}
  \centering
    \begin{tabular}{|c|c|c|}
    \hline 
\multicolumn{1}{|c|}{\multirow{2}{*}{$N_1$} }                                       & \multicolumn{2}{c|}{Unstable $2$-partitions}            \\ \cline{2-3}
 \multicolumn{1}{|c|}{} &     Proportional & Shapley   \\ \hline
         $2-9$  & None & None \\ \hline
         $10-17$ &  $w \in \{14,\cdots,21\}$  &   None   \\ \hline
         $18-40$ & $w \in \{20,\cdots,44\}$ & None \\ \hline
    \end{tabular}
    \vspace{6mm}
    \captionof{table}{Unstable partitions under RB-PA for different \\ allocation rules with~$N_2=N_3=N_4=2, \Lambda = 13$}
    \label{tab:my_label}
\end{minipage}
\begin{minipage}{.45\textwidth}
     \centering     \includegraphics[trim = {3cm 8cm 0cm 8cm}, clip, scale = 0.5]{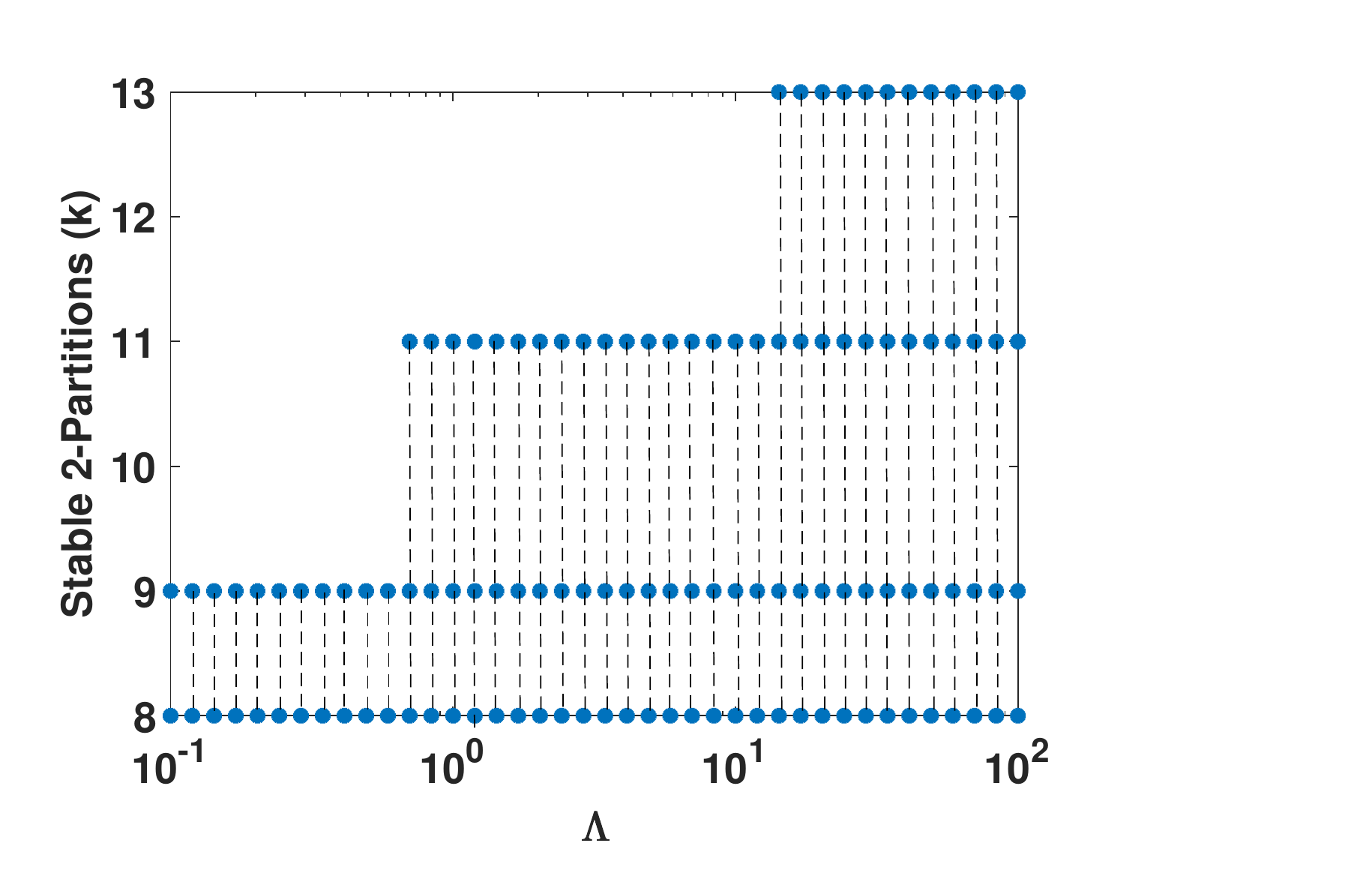}
     \captionof{figure}{Set of stable partitions (under RB-IA) v/s~$\Lambda$  (on log scale) for~$N_1 = 7, N_2 =  N_3 = N_4 = N_5 = 2 $}
     \label{fig:stable partitions}
\end{minipage}
\end{figure}

 \textbf{RB-PA rule:} Next, we study the RB-PA rule. Our aim is to first compare the stability of two allocation mechanisms---proportional allocation and Shapley value. Consider the following example with~$\Lambda = 13$ and~$4$ service providers. Here~$N_1$ is varied from~$2-41$, while  the remaining service capacities are fixed at~$N_2 = N_3 = N_4= 2$. 
Table~\ref{tab:my_label} presents the set of $2$-partitions that are unstable under each allocation mechanism. Here,~$w$ denotes the number of servers in the coalition that includes provider~$1$. For example, the second row considers the cases where~$N_1$ lies between~$10$ and~$17$. In all these cases, the proportional payoff vector renders   those~$2$-partitions with~$w \in \{14, 15, \cdots, 21\}$ unstable, whereas all two partitions are stable under Shapley value. This suggests that Shapley value renders more partitions stable in comparison to the  proportional payoff vector.

We consider another such example with 3 agents,~$\Lambda=100,$~$ N_1=80,$~$ N_2=20$ and~$N_3=5$. By Theorem~\ref{Thm_SV_small}.$(i)$,~$(\P,\Phi_s^\P)$ is stable for~$\P = \{\{1,2\},\{3\}\}$. However, we find (numerically) that~$(\P, \Phi_p^\P)$ is not stable (it is blocked by~$Q = \{1\}$). (Numerically, we find that~$k^*=\{80\}$, implying~$\P$ does not satisfy the hypothesis of Theorem~\ref{Thm_R_rule_stability}, as expected.) 


\textbf{Impact of congestion: }In Figure~\ref{fig:stable partitions}, we consider a final example that demonstrates how the set of stable partitions under RB-IA varies with the market size~$\Lambda.$ Here, we consider five service providers with service capacities~$N_1 = 7, N_2= N_3 = N_4 = N_5 = 2.$ 
Note that the left and right extremes in the figure are consistent with the light-traffic and heavy traffic results (Theorems~\ref{low_traffic} and~\ref{Thm_heavy} respectively). In particular, in light traffic, the only stable duopolies are those that are nearly matched with respect to service capacity---one where the dominant coalition is composed of agents~$2, 3, 4,$ and~$5$ ($k = 8$) and another the dominant coalition is composed of agent~$1$ and one of the remaining agents ($k=9$). In heavy traffic, all duopolies are stable. Importantly, the figure shows that the set of stable duopolies grows monotonically with~$\Lambda.$


 %
 

\section{Conclusions and Future Work}


\new{Our work highlights that in competitive service systems enjoying statistical
economies of scale, coalition formation games have very distinct
equilibria when the total payoff across agents is a constant. In
particular, we demonstrate that \emph{duopolies} emerge, with the
dominant coalition exploiting economies of scale to corner a
disproportionate fraction of the total payoff.

This work motivates future work along several directions.}{} Firstly, one could explore alternative models for a coalition's utility. For instance, one could define the utility of a coalition to be the rate of customers \emph{served} (rather than the rate of customer \emph{arrivals}); this is meaningful in scenarios where providers only earn revenue when a customer is successfully served. Preliminary analysis suggests that this modification of the utility structure alters the nature of stable equilibria. More generally, this work motivates a systematic understanding of how payoff structures influence
the nature of equilibria in partition form games. 

Another potential direction of inquiry involves exploring the effect of different queueing models, including models where customers can wait for service with/without balking or reneging.

Finally, it would also be interesting to explore dynamic variants of coalition formation games. This would entail examining whether any limiting behaviors emerge (particularly when stable equilibria do not exist). 

\begin{APPENDICES}
\section{Characteristic Form games }
\label{appendix_A}

\new{A game in characteristic form~\cite{aumann1961} can be defined using  the tuple,~$(\mathcal{N},  \nu,\mathcal{H})$, 
where: 
(a)~$\mathcal{N}$ denotes the set of~$n$ agents; 
(b)~$\nu$ is called a characteristic function and  for any~$C \subseteq \mathcal{N}$,~$\nu(C)$  denotes the set  of all possible
 payoff vectors of dimension~$n$ that agents in~$C$ can jointly achieve;  
 and
(c)~$\mathcal{H}$ is the set of all possible payoff vectors of dimension~$n$ (such vectors are also referred to as allocation vectors in literature), which are achievable.}{} We say~$(\mathcal{N},  \nu,\mathcal{H})$ is an \textit{ordinary game} (see~\cite{aumann1961}) if:~$\x \in \nu(\mathcal{N})$ if and only if there is a~$\textbf{y} \in \mathcal{H}$ such that~$x_i \leq y_i  \text{ for all }  i.$ 

\new{In this appendix, we provide the details of how our problem can be recast as a characteristic game.  
Let~${\cal F} (\P) $ be the set of all feasible payoff vectors
under partition~$\P$,  these are the vectors that satisfy the following:     the sum of payoffs of all  agents in any coalition~$S$ is less than or equal to that obtained by~$S$ under partition~$\mathcal{P}$ at WE,~$\lambda_S^\mathcal{P}$. Hence  
\begin{equation}
{\cal F} (\P) := \left \{\x = [\xs_i] : \sum_{i \in S} \xs_i \leq \lambda_S^\mathcal{P} \text{ for all }  S \in \P   \right \}.
\end{equation}
Thus $\mathcal{H}$, the set of all achievable/feasible payoff vectors is~$
\mathcal{H} = \cup_{\mathcal{P}} {\cal F} (\P) .
$}{}
Observe that for grand coalition,~$\mathcal{F}(\mathcal{N}) = \mathcal{H}$ and hence is convex. We are now left to define the characteristic function~$\nu$.

\new{{\bf  Characteristic function using pessimal rule:}}{}
\new{The  characteristic function precisely describes the set of all possible divisions of the anticipated worth  of any coalition.}{} One can define such a function for partition form games using an appropriate anticipation rule~\cite{pessimistic}. \new{There are many known anticipatory rules to define characteristic function}{}, also described in Section~\ref{sec:classical}. 

According to the most widely used \new{\textit{pessimistic anticipation rule}~\cite{pessimistic}, the agents in deviating coalition~$C$ assume that the outside agents arrange themselves to hurt the agents in~$C$ the most. 
Further, the minimum utility that coalition~$C$ can achieve irrespective of the arrangement of the agents outside this coalition is given  by~$
{\underline \nu}_C   := \min_{ \P : C \in \P } \max_{\x \in {\cal F} (\P)}  \sum_{i \in C}  \xs_i  
$
(observe in our case,~${\underline \nu}_C = \ulam_C$).
Thus, the characteristic function~$\{\nu(C); \text{ for all } C\}$ under pessimal rule is given by the following: for any coalition~$C$,~$
\nu(C) = \left \{\x : \sum_{i \in C}  \x_i \leq \underline{\nu}_C \right \},
$ is  the set of possible payoff vectors that agents in~$C$ can jointly achieve independent of the arrangement of outside agents. }{}
From the above definition, it is clear that our game is an ordinary game. 

\textbf{Stability:} \new{To study the stability aspects, one needs to understand if a certain coalition can `block' any payoff vector.
Blocking by a coalition implies that coalition is working as an independent unit and has an anticipation of the value it can achieve (e.g., irrespective of arrangements of others under pessimal rule). 
If the division of this anticipated value among the members of the coalition, under any given allocation rule,  renders the members to achieve more than that in the current payoff vector then the coalition has tendency to oppose the current arrangement or the payoff vector. 

\textit{Blocking:} 
A payoff vector~$\x \in \mathcal{H}$ is blocked by a coalition~$C$ if there exist a payoff vector~$\textbf{y} \in \nu(C)$ such that~$
y_i > \xs_i \text{ for all } i \in C.
$

With these definitions in place, we now give define a related solution concept called R-core, which is an extension of the classical definition of \textit{core}, for transferable utility games (in non-partition form games). 

\textbf{R-core~\cite[Section 3]{aumann1961}:} 
We define R-core~$\mathscr{C}(\mathcal{H})$ to be the set of vectors in~$\mathcal{H}$ which cannot be blocked by any other member of~$\mathcal{H}$. }{}

The authors in~\cite{hafalir} studied the properties of this core under the name \textit{c-core} (which is also popular by the name $\alpha$-core in literature). In~\cite[Corollary 2]{hafalir}, they showed that a convex partition form game necessarily has a non-empty core. However, one can easily check that our game is not convex as in~\cite{hafalir} and hence, it is not clear if core is non-empty or not. In fact, in Theorem~\ref{thm_impossible}, we showed that the R-core is empty for our game. We hence introduce more generalised and relevant notions of stability in this paper.

\section{Proof of Theorem~\ref{Thm_WE}}
\label{appendix_B}

\textbf{ Proof of  Existence and Uniqueness:}
Let the size of a partition be denoted by~$p$.
The first step of this proof is to show the existence and uniqueness of WE for the case when~$p=2$.  In the next step,  using induction we prove the existence for any general~$p=m>2$ using the corresponding results for~$m-1$. In the third step we show the continuity of the WE, to be precise the arrival rates at WE for~$m$. The last step attributes to the uniqueness of our solution.

\textit{Step 1: Existence and Uniqueness of WE for~$p=2$} 

To obtain WE, the following equation needs to be solved:~$
B_{C_1}^\P(N_{C_1},a_{C_1}^\P) = B_{C_2}^\P(N_{C_2},a_{C_2}^\P).
$
Define a function~$f:= B_{C_1}^\P(N_{C_1},a_{C_1}^\P) - B_{C_2}^\P(N_{C_2},a_{C_2}^\P)$. Then,~$f$ is a function of~$\lambda_{C_1}^\mathcal{P} \in [0,\Lambda]$ since~$\lambda_{C_2}^\mathcal{P}= \Lambda-\lambda_{C_1}^\mathcal{P}$. 
\begin{itemize}
	\item 
	At~$\lambda_{C_1}^\mathcal{P}=0$ we have~$B_{C_1}^\P(N_{C_1},a_{C_1}^\P)=0$ and~$B_{C_2}^\P(N_{C_2},a_{C_2}^\P) >0$, thus~$f(0) <0$.
	
	\item At~$\lambda_{C_1}^\mathcal{P}=\Lambda$ we have~$B_{C_1}^\P(N_{C_1},a_{C_1}^\P)>0$ and~$B_{C_2}^\P(N_{C_2},a_{C_2}^\P) =0$, thus~$f(\Lambda) >0$. 
\end{itemize}
Then,~$B_{C_1}^\P(N_{C_1},a_{C_1}^\P)$ and~$B_{C_2}^\P(N_{C_2},a_{C_2}^\P)$ are polynomial functions with denominator~$> 1$ and hence are continuous functions. This implies that~$f$ is a continuous function.

Thus,~$f$ satisfies the hypothesis of Intermediate Value Theorem (IVT). Using IVT, there exists a value of~$\lambda_{C_1}^\mathcal{P} = \lambda^* \in (0,\Lambda)$ such that~$f(\lambda^*)=0$.
The uniqueness of~$\lambda^*$ follows since~$B_{C_1}^\P(N_{C_1},a_{C_1}^\P)$ and~$B_{C_2}^\P(N_{C_2},a_{C_2}^\P)$ are strict increasing functions of~$\lambda_{C_1}^\mathcal{P}$ and~$\lambda_{C_2}^\mathcal{P}$ respectively. 

 \textit{Step 2: Existence for general~$p = m > 2$}
 
 To prove the existence for any general~$m>2$, we assume that a unique WE exists for~$p=m-1$, i.e.,~$\lambda_{C_1}^{\P}, \cdots, \lambda_{C_{m-1}}^{\P}$ with corresponding common blocking probability~$B^*$.
 
 With~$m$ units we can initially fix~$\lambda_{C_{m}}^{\P} = 0$ and obtain WE corresponding to the remaining units, which we have assumed to exist. With increase in~$\lambda_{C_m}^{\P}$, $\Lambda - \lambda_{C_m}^{\P}$ which is the total share of remaining agents, decreases. From part $(i)$ of this theorem  applied to the case with~$m-1$, we know that the corresponding WE solution for these agents also decreases. This implies that the common blocking probability for~$C_1, \cdots, C_{m-1}$ reduces while blocking probability of~$C_m$ increases (see~\eqref{Eqn_PB}). Using similar arguments as above and treating~$C_1, \cdots, C_{m-1}$ as one while defining function for IVT (continuity is obtained from Step $3$, with~$m-1$), one can show that WE exists.

\textit{Step 3: Continuity of Optimisers, i.e., WE:} 
Consider the following function~$g$ for~$m$ coalitions in partition~$\P$:~$
g(\Lambda,{\bm \lambda}) := \sum_{C_j \in \mathcal{P}; 1<j\leq m} (B_{C_1}^\P-B_{C_j}^\P)^2,
$
where~${\bm \lambda}$ is the vector of arrival rates for all~$C_j \in \P$. Then, we define~$g^*(\Lambda,{\bm \lambda}^*) = \min_{\{{\bm \lambda}: \sum_j \lambda_j = \Lambda \} }g(\Lambda,{\bm \lambda}) $.
Observe that the (unique) minimizer~${\bm \lambda}^*$ of the function~$g$ is the (unique) WE for our queueing model, and that the function~$g$ is jointly continuous. Thus, using Maximum Theorem we have that~$g^*$ and~${\bm \lambda}^*$ is continuous in~$\Lambda$.

\textit{Step 4: Uniqueness of WE }
To prove the uniqueness of the WE, we assume the contradiction,  i.e., say~$(\lambda_1,\cdots,\lambda_m)$ and~$(\lambda_1',\cdots,\lambda_m')$ are two distinct WEs. One can have the following cases: 

 \textit{Case 1: There exist multiple WEs with same common blocking probability~$B^*$} 
  This implies that some of the units in partition are obtaining different arrival rates in the multiple WEs such that they have common~$B^*$, i.e., say $\lambda_i' \ne \lambda_i$. However, this is not possible since blocking probability is a strictly increasing function of arrival rate. 
 
 \textit{Case 2: There exist multiple WEs with different common blocking probability~$B^*$ and~${\hat B}^*$} 
 
 Without loss of generality, we can assume that~$B^* < \hat{B}^*$. This implies that the arrival rates to the units with common blocking probability~$\hat{B}^*$ is more (since blocking probability is an increasing function of arrival rate). However, the total arrival rate is fixed at~$\Lambda$ which implies that one of the WE does not satisfy~$\sum_{C_j \in \P} \lambda_{C_j}^{\P} = \Lambda$. \Halmos

\textbf{Proof of  All units used}  
For contradiction, let us assume that the customers split themselves amongst some strict subset of units of partition~$\P$. Then, each unit with zero arrivals have a zero blocking probability while units with non-zero arrivals have some strict positive blocking probability. However, this contradicts the fact that the coalitions having zero arrivals should have a higher blocking probability than others at WE.

Hence at WE, each of the units in partition~$\P$ obtain non-zero arrival rates. \Halmos

\textbf{Proof of part (i)}  Let~$\lambda_{C_1}^\P, \cdots, \lambda_{C_k}^\P$ be the individual arrival rates corresponding to partition~$\P$ at WE  (satisfies~\eqref{Eqn_WE_properties}) for the coalitions~$C_1, \cdots, C_k$ respectively with the total arrival rate~$\Lambda>0$. Let the   corresponding  common  blocking probability be~$B^*$. When the total arrival rate is increased to~$\Lambda'$, the individual arrival rates to the providers at WE are changed to~$\lambda_{C_1}^{'\P},\cdots, \lambda_{C_k}^{'\P}$ and the   corresponding  common  blocking probability is changed to ~$\hat{B}^*$. Note that these splits to the individual operating units must satisfy:
\begin{equation}
\sum_{i=1}^k \lambda_{C_i}^{'\P} = \Lambda' > \Lambda =  \sum_{i=1}^k \lambda_{C_i}^{\P}  \text{ and }  \text{ for any partition } \P.
\label{total_conserved}
\end{equation} 

\noindent
Next we will show that~$\lambda_{C_j}^{'\P} \leq \lambda_{C_j}^{\P}$ is not possible for any~$C_j \in \P$. Using~\eqref{total_conserved}, we know that at least one of the units have higher individual arrival rates at new WE, i.e,~$
\lambda_{C_j}^{'\P} > \lambda_{C_j}^{\P} \text{ for at least one } C_j \in \P.
$
This means that the common blocking probability at new WE is increased, i.e.,~$\hat{B}^* > B^*$. Now since blocking probability is a strictly increasing function of arrival rates, we have that arrival rate to each coalition is increased at new WE for~$\Lambda'$, i.e.,~$\lambda_{C_j}^{'\P} > \lambda_{C_j}^{\P}$ for all~$C_j \in \P$. 

Hence, WE is an increasing function of  $\Lambda$. \Halmos

\textbf{Proof of part (ii)} Let~$\lambda_{C_1}^\P, \cdots, \lambda_{C_k}^\P$ be the individual arrival rates corresponding to partition~$\P$ at WE for the coalitions~$C_1, \cdots, C_k$ respectively. 
 Let the   corresponding  common  blocking probability be~$B^*$. Observe that the blocking probability of~$C_i$ and~$C_j$ units also equals~$B^*$, and hence the merger~$M = C_i \cup C_j \neq \mathcal{N}$ has strictly smaller  blocking probability, i.e.,~$B_M < B^*$,  if the joint arrival rate was~$\lambda_{C_i}^\P + \lambda_{C_j}^\P$. 
From~\eqref{Eqn_PB} the  blocking probability is a strictly increasing function of arrival rate. Thus the new WE after merger is formed with a (strict) bigger arrival rate to the merger, as again at the new WE the new blocking probabilities of all coalitions~$C \in {\P}'$ should be equal by~\eqref{Eqn_WE_properties}.  \Halmos

\textbf{Proof of  part  (iii)} 
 Consider a system with identical servers. We know that when any number of identical servers combine with their arrival rates, the combined blocking probability reduces. This reduction is more when the number of servers combining are more, i.e.,
\begin{equation}
B(N,a) > B(LN,La) > B(MN,Ma).
\label{Eqn_common_result}
\end{equation}
where~$0<L< M$ are constants,~$B$ is the blocking probability,~$N$ is the number of servers and~$a$ is the offered load.
Now if we consider that the coalition with~$N_{C_1}$ and~$N_{C_2}$ servers gets exactly~$N_{C_1}/N$ and~$N_{C_2}/N$ share of total arrival rate~$\Lambda$ at WE respectively.
Using~\eqref{Eqn_common_result}, we have that coalition with~$N_{C_1}$ servers has strictly smaller blocking probability. From~\eqref{Eqn_WE_properties}, the blocking probability of each unit at WE is same. So, the arrival rate to coalition with~$N_{C_1}$ and~$N_{C_2}$ servers need to be increased and reduced respectively to achieve the WE. 

Hence, coalition with~$N_{C_1}$ and~$N_{C_2}$ servers satisfy~$ 
\frac{\lambda_{C_1}^{\P}}{N_{C_1}} > \frac{\Lambda}{N} > \frac{\lambda_{C_2}^{\P}}{N_{C_2}}.
$ \Halmos

\section{Rest of the proofs}
\label{appendix_C}

\proof{Proof of Theorem~\ref{thm_impossible}:} 
 Consider any  configuration, say~$(\P,\Phi)$. 
From~\eqref{eq:blocking_PA}, the configuration is stable if and only if
\begin{equation}
\label{Eqn_reqd_conditions}
\sum_{i \in C} \phi_i \ge \underline{\lambda}_C \text{ for all } C \notin \P \text{ and } C \subset \mathcal{N}. 
\end{equation} 

\textit{Case 1: All players are alone in~$\P$}

In such a case, for some player~$j$, consider the merger coalition~$M = \mathcal{N}-\{j\}$. Then from Theorem~\ref{Thm_WE}.$(ii)$,
\begin{equation*}
 \ulam_M  > \sum_{C_l \in M} \lambda_{C_l}^ \mathcal{P} = \sum_{i\ne j} \phi_i,
\end{equation*}
which implies that~$M$ blocks the prevalent configuration under the GB-PA rule.

\textit{Case 2: There exists at least one coalition~$C \in \P$ such that~$|C| \ge 2$}

This implies that~$S_a = \mathcal{N}-\{a\} \notin \P$ for all~$a \in C$. We will show that some~$a \in C,$ either~$S_a$ or~$\{a\}$ will block the prevailing configuration. 

\noindent
\textit{Case 2(a): The configuration is blocked by~$S_a$ for some~$a \in C.$} In this case, the instability of the coalition follows immediately.

\noindent
\textit{Case 2(b): The configuration is not blocked by~$S_a$ for any~$a \in C.$} In this case,
$$
\sum_{i \in S_a} \phi_i \ge \underline{\lambda}_{S_a} = \Lambda - \ulam_{\{a\}}
$$ for all~$a \in C.$ This is equivalent to the statement~$\phi_a \leq \ulam_{\{a\}}$ for all~$a \in C.$
However, there exists a ${\hat a} \in C$ such that~$\phi_{\hat a} < \ulam_{\{{\hat a}\}}$ since~$\sum_{q \in C} \phi_i = \sum_{q \in C} \ulam_{\{q\}} \le \sum_{q \in C} \lambda_{\{q\}}^{\P'}  < \lambda_C^\P$. Thus, the configuration $(\P, \Phi)$  is blocked by~$\{{\hat a}\}.$ \Halmos 
\endproof


\proof{Proof of Theorem~\ref{Thm_duo_mono}:}
Consider a partition~$\mathcal{P} = \{C_1,C_2,\cdots, C_k\}$ with cardinality greater than~$2$. Let~$M$ be the merger coalition containing all coalitions of~$\P$ except one, i.e.,~$
M = \cup_{i=2}^k C_i \text{ and } \P' = \{C_1,M\}.
$ Then from Theorem~\ref{Thm_WE}.$(ii)$,~$
 \lambda_{M}^{\mathcal{P}'} = \ulam_M^{\P'}  > \sum_{C_i \in M} \lambda_{C_i}^ \mathcal{P},
$
which is same as the condition required for blocking by mergers under RB-IA rule. 

 Hence, there exists a configuration/payoff vector such that each of the members in $M$ obtain strictly better and thus, such a partition is not stable.\Halmos
 \endproof

\proof{Proof of Theorem~\ref{Thm_GC}:} There can be no merger from~$\P_G$,  and we only need to check if an appropriate split can block a configuration~$(\P_G, \Phi)$, under consideration.

\begin{enumerate}[(i)]
    \item \textit{When~$N_1 < \sum_{i \in \mathcal{N}; i \neq 1}N_i$}

\begin{enumerate}[(a)]
\item We first consider payoff vectors~$\Phi$ that satisfy 
\begin{equation}
   \sum_{i=2}^n \phi_i < \left (   1-\frac{N_1}{N} \right ) \Lambda.
   \label{Eqn_payoff1}
\end{equation} 
Let~$S :=\{2, 3, \cdots, n\}$  be the coalition  made of all agents except agent $1$. We will prove that this  coalition will block the configuration of the form stated above.

 Since coalition~$S$ has more than $N/2$ servers, it must satisfy  the following (from Theorem~\ref{Thm_WE}.$(iii)$):~$
\lambda_S^{\P'} =  \underline{\lambda}_S > \Lambda \left (  1-\frac{ N_1}{N} \right )    ,  \mbox{ where }   {\P'}  := \{S, \{1\} \}, 
$
which is same as~\eqref{Eqn_condition_S}.
Further, from~\eqref{Eqn_payoff1},~$
\lambda_S^{\P'} > \Lambda \left (  1-\frac{ N_1}{N} \right ) > \sum_{i=2}^n \phi_i, 
$
which implies that~\eqref{Eqn_condition_S_pt2} is also satisfied by coalition~$S$.

 Hence,~$(\P_G,\Phi)$ is blocked by coalition~$S$.

\item Next, we consider payoff vectors that satisfy \begin{eqnarray}
\sum_{i=2}^n \phi_i \geq  \left (  1-\frac{N_1}{N}  \right ) \Lambda. \label{Eqn_condition1_new}
\end{eqnarray}

Suppose, for the sake of obtaining a contradiction, that,~$\{\P_G,\Phi\}$ is stable. Since~$N_1$ is the agent with  maximum number of servers,~$S_k := S \backslash \{k\} \cup \{1\}$ has~$N_{S_k}  > N/2$ for any~$k \ge 2$. By Theorem~\ref{Thm_WE}.$(iii)$ such coalitions satisfy~\eqref{Eqn_condition_S}. Thus, the stability of~$\{\P_G,\Phi\}$ implies that~\eqref{Eqn_condition_S_pt2} must be violated for the same coalitions. That is, we have~$
\sum_{i \in S_k} \phi_i \ge \ulam_{S_k}  >  \frac{\sum_{i \in S_k} N_i}{N}\Lambda,  \mbox{ for each }  k > 1, 
$ in view of Theorem~\ref{Thm_WE}.$(iii)$.
By adding all the above inequalities with~$k=2, \cdots, n$, we have:
\begin{eqnarray*}
(n-1)\phi_1+(n-2) \sum_{i=2}^n \phi_i   >   \frac{(n-1)N_1+(n-2)\sum_{i=2}^n N_i}{N} \Lambda,
\end{eqnarray*}
which implies,
 \begin{eqnarray*}
 \phi_1 + (n-2)\Lambda > \Bigg(\frac{N_1+(n-2)N }{N} \Bigg)\Lambda = \frac{N_1}{N}\Lambda + (n-2)\Lambda, 
\end{eqnarray*}
since~$\sum_{i=1}^n \phi_i = \Lambda,   \sum_{i=1}^n N_i = N.$
Thus we have,~$
\phi_1    >  \frac{N_1}{N}\Lambda $ which contradicts~\eqref{Eqn_condition1_new}. Thus,~$(\P_G,\Phi)$ is unstable under RB-IA rule.
\end{enumerate}

\item \textit{When~$N_1 \ge \sum_{i \in \mathcal{N}; i \neq 1}N_i$}

In this case, the coalitions that satisfy condition~\eqref{Eqn_condition_S} for blocking under RB-IA are exactly those coalitions that contain player~$1$ (from Theorem \ref{Thm_WE}.$(iii)$). However, for any such coalition, the condition~\eqref{Eqn_condition_S_pt2} for blocking under RB-IA gets violated so long as~$
\phi_1 \ge \max_{C} \ulam_C \text{ for all } C \subset \mathcal{N} \text{ containing agent 1.}
$
Thus, any allocation~$\Phi$ satisfying the above bound on~$\phi_1$ is guaranteed to be stable under RB-IA. \Halmos
\end{enumerate}
\endproof

\proof{Proof of Theorem \ref{Thm_two_partition}:}
Part~$(i)$ follows from part~$(ii)$, proved below,  as~$k^*$ exists.

$(ii)$ Any~$2$-partition~$\P = \{C_1,C_2\}$ cannot be blocked by mergers since merger lead to~$\P_G$ and~\eqref{Eqn_condition_M} is not satisfied. Next we look at splits. Say~$C_1 \in \mathbb{C}^*$. Then it follows from the definition of~$\mathbb{C}^*$ that there exists no coalition~$C \subset C_1$ such that it satisfies~\eqref{Eqn_condition_S}. Further, coalition~$C_2$ cannot do better by splitting.
Hence, any partition with one of the coalitions belonging to~$\mathbb{C}^*$ is a stable partition under RB-IA rule.

$(iii)$ Once again, it is easy to verify that a merger cannot block any~$2$-partition~$\P.$ Next, we check for splits.
    Any split leads to a coalition with a number of servers less than~$N/2$, and hence from Theorem~\ref{Thm_WE}.$(iii)$,~\eqref{Eqn_condition_S} is not satisfied and hence, no split is feasible. Thus,~$\P$ is stable under RB-IA rule. \Halmos
    \endproof

\proof{Proof of Theorem \ref{Thm_R_rule_stability}: } 
$(i)$ Consider any configuration~$(\P_G, \Phi )$  with GC.
The proof of this part can be split into two cases:

\textit{Case 1: When~$N_1 < \sum_{i \in S; i \neq 1}N_i \text{ for some } S \subset \mathcal{N}$}

Under RB-PA rule for the configuration to be stable, we need to ensure that the  following system of equations are satisfied simultaneously.
\begin{eqnarray}
\sum_{i \in C} \phi_i \geq \ulam_C \text{ for all } C \subset \mathcal{N} \text{ and, }
\sum_{i \in \mathcal{N}} \phi_i  = \Lambda.  \label{Eqn_conditions}
\end{eqnarray} 
However, a subset of these equations itself admit no feasible solution (as proved in Theorem~\ref{Thm_GC}).   Thus, such a system of equations does not have a solution and  hence~$(\P_G, \Phi )$ is unstable for any payoff vector $\Phi$.

\textit{Case 2: When~$N_1 \ge \sum_{i \in \mathcal{N}; i \neq 1}N_i$}

Once again we need to satisfy~\eqref{Eqn_conditions} to prove that~$(\P_G, \Phi)$ is stable. In particular those equations will also have to be satisfied  for subsets~$S$ such that~$|S| = n-1$.
If there exists a payoff vector~$\Phi$  that satisfies all such conditions,  consider one such~$S$ and say~$j \notin S$.  Then  from~\eqref{Eqn_conditions},~$\phi_j = \Lambda - \sum_{i \in S} \phi_i  \leq  \Lambda -\ulam_S 
= \ulam_{\{j\} }.$
If~$\phi_j < \ulam_{\{j\}}$ for some~$j$ then configuration~$(\P_G, \Phi)$  is blocked by~$\{j\}$ under RB-PA rule. Otherwise if~$\phi_j = \ulam_{\{j\}}$ for all~$j \in \mathcal{N}$ then~$\sum_{i \in \mathcal{N}} \phi_i = \sum_{i \in \mathcal{N}} \ulam_{\{j\}} < \Lambda$ and thus~\eqref{Eqn_conditions} is not satisfied.
Hence~$(\P_G, \Phi)$ is unstable for any payoff~$\Phi$. 

$(ii)$ Since the condition required for a merger to be successful under RB-PA rule is same as under RB-IA rule, the result follows from Theorem~\ref{Thm_duo_mono}.

$(iii)$ 
%
When the  payoff vector is given by equation~\eqref{Eqn_PSA}, the RB-PA and RB-IA rules 
are equivalent to each other.  
Thus, the result follows from Theorem~\ref{Thm_two_partition}.


 Moreover because of the continuity of~$\Phi$, we have the next result. \Halmos
  \endproof
	 
	 \proof{Proof of Theorem \ref{Thm_SV_small}:} Consider any~$2$-partition~$\P = \{C_1,C_2\}$.
	 
$(i)$ W.l.o.g., say coalition~$C_1 = \{i,j\} $.
	 From~\eqref{Eqn_SV} and~\eqref{eq_subcoalition_worth}, the share of player~$i$ is given by:
	 \begin{eqnarray}
 \phi_i  & = & \frac{1}{2} \left( \lambda_{C_1}^\P - \lambda_{\{j\}}^{\P'} \right) + \frac{1}{2} \lambda_{\{i\}}^{\P'} \ > \lambda_{\{i\}}^{\P'} > \ulam_{\{i\}}, \text{ where } \P' = \{\{i\},\{j\},C_2\}.\nonumber 
  \end{eqnarray}
   The first inequality holds since~$\lambda_{C_1}^\P > \lambda_{\{i\}}^{\P'}+\lambda_{\{j\}}^{\P'},$ and the second follows from Theorem~\ref{Thm_WE}.$(ii)$. Thus, a split of~$C_1$ does not block the configuration~$(\P, \Phi_s^\P)$. Further, a merger cannot block the configuration due to the constant sum nature of the game.
  
An identical argument also applies for part $(ii)$. \Halmos
	 \endproof

\proof{Proof of Theorem \ref{Thm_heavy}:} Consider any~$2$-partition~$\P = \{C_1,C_2\}$ with~$k := N_{C_1} \ge N_{C_2}$ It is easy to see that the~$2$-partition cannot be blocked by a merger under RB-IA/RB-PA rules. It therefore suffices to check for stability against splits.
\begin{enumerate}[(i)]
    \item Be relaxing $k$ to be a real number, we show that $\Psi$ is increasing for all $k$ in Lemma~\ref{lem_psi_inc}. Using Lemma~\ref{lem_psi_inc},  no split satisfies~\eqref{Eqn_condition_S} and hence~$\P$ is stable.
    \item Under the proportional payoff vector~$\phi_p^\P$,~\eqref{eq_def_RBPA} is equivalent to~\eqref{Eqn_condition_S}, and hence the result follows.
\end{enumerate}
Below, we prove Lemma~\ref{lem_psi_inc}.
\Halmos
\endproof

\begin{lemma}
\label{lem_psi_inc}
Consider any $\epsilon>0$.  Then there exists a $\bar{\Lambda}$ such that for all $\Lambda \ge {\bar \Lambda},$ $\Psi:=\lambda_1/k$ is strictly increasing in $k$ over $N/2 \le k \le N-\epsilon$. 
\end{lemma}
\proof{Proof:}
To prove this result, we work with the analytical extension of the Erlang-B formula (see~\cite{jagerman}) so that~$k$ may be treated as a real number. Under this extension, it is easy to see that the Wardrop splits are uniquely defined for real-valued service capacities. For any 2-partition, differentiating~$\Psi$ with respect to~$k$, we have
$
    \frac{d \Psi}{d k}  =  \frac{d}{dk} \left(\frac{\lambda_1}{k} \right) \ = \ \frac{1}{k} \left(\frac{d \lambda_1}{d k} - \frac{\lambda_1}{ k} \right). \nonumber
$
Thus, to prove the theorem, it suffices to show that given~$\epsilon > 0$, there exists a~$\bar{\Lambda}$ such that for any~$\Lambda \geq \bar{\Lambda}$,
\begin{equation}
    \frac{d \lambda_1}{d k} - \frac{\lambda_1}{ k} > 0 \text{ for all } k \in \left[\frac{N}{2}+\epsilon,N-\epsilon \right]. \nonumber
\end{equation}
Towards this, we know that the arrival rates at WE~($\lambda_1$) are obtained by equating the blocking probabilities of the two coalitions. The reciprocal of the blocking probability of a coalition with~$k$ servers and offered load~$a$ admits the following integral representation (see~\cite{jagerman}):
\begin{equation}
    R(k,a) = a\int_0^\infty  h(t; a, k) dt
    \mbox{ where }
   h(t; a, k) =     (1+t)^k e^{-a t} . \nonumber
\end{equation}
Thus, the WE satisfies~$
    R(k,\lambda_1) - R(N-k,\Lambda-\lambda_1) = 0, \nonumber 
$
    which is equivalent to
   \begin{eqnarray}
    \lambda_1 \int_0^\infty h(t;\lambda_1,k) dt - (\Lambda-\lambda_1)\int_0^\infty h(t;\Lambda-\lambda_1,N-k) dt = 0.  
    \label{eq_integral_pb}
\end{eqnarray}
 Differentiating both sides of the above  with respect to~$k$ using Lemma~\ref{lem_interchange}\Long{ and rearranging:}{, we obtain~\eqref{Eqn_Integral}.

 \vspace{-4mm}
 {\small
 \begin{eqnarray}
    \frac{d \lambda_1}{dk} \left[\int_0^\infty(1+t)^k e^{-\lambda_1 t} dt \right] + \lambda_1 \left[ \int_0^\infty(1+t)^k ln(1+t)e^{-\lambda_1 t} dt - \int_0^\infty(1+t)^k e^{-\lambda_1 t} t \left(\frac{d \lambda_1}{dk} \right) dt\right] + \frac{d \lambda_1}{dk} \left[\int_0^\infty(1+t)^{N-k} e^{-(\Lambda-\lambda_1) t} dt \right] \hspace{-7mm} & \nonumber \\
& \hspace{-140mm}  - (\Lambda-\lambda_1) \left[-\int_0^\infty(1+t)^{N-k} ln(1+t)e^{-(\Lambda-\lambda_1) t} dt + \int_0^\infty(1+t)^{N-k} e^{-(\Lambda-\lambda_1) t} t \left(\frac{d \lambda_1}{dk} \right) dt \right] = 0. \nonumber \label{Eqn_Integral}
\end{eqnarray}}
Rearranging the terms in~\eqref{Eqn_Integral} we obtain~\eqref{Eqn_derivlam1}.}

\vspace{-8mm}
{\small \begin{eqnarray}
    \frac{d \lambda_1}{d k}  =  \frac{-\lambda_1 \int_0^\infty h(t;\lambda_1,k) \ln(1+t)  dt - (\Lambda-\lambda_1 )\int_0^\infty h(t;\Lambda-\lambda_1,N-k) \ln(1+t) dt}{ \int_0^\infty h(t;\lambda_1,k) dt + \int_0^\infty h(t;\Lambda-\lambda_1,N-k) dt-\lambda_1 \int_0^\infty h(t;\lambda_1,k)t dt - (\Lambda-\lambda_1 )\int_0^\infty h(t;\Lambda-\lambda_1,N-k)t dt}.\nonumber \label{Eqn_derivlam1}
\end{eqnarray}}
Observe that each of the integrals in the above expression is of the form~$\int_0^\infty f(t;k)e^{-\lambda_1t} dt \text{ or } \int_0^\infty f(t;k)e^{-(\Lambda-\lambda_1)t} dt.$
In heavy traffic, since~$\lambda_1$  and~$\Lambda-\lambda_1$ tend to infinity (see Lemma~\ref{lem_accuracy_heavy} below) the value of these integrals is dominated by the behavior of the integrand around zero. Accordingly, one can approximate these integrals using a Taylor expansion of~$f(t)$ around~$t=0.$ Formally, using Lemma~\ref{lem_error_taylor} below (it is easy to show that all the integrals above satisfy the hypotheses of Lemma~\ref{lem_error_taylor}), we have~$\frac{d \lambda_1}{dk} =- \frac{T_1}{T_2}, \text{ where }$ 

\vspace{-9mm}
{\small \begin{align*}
T_1 &= \lambda_1\left(\frac{1}{\lambda_1^2}+\frac{2k-1}{\lambda_1^3}+\frac{3k^2-6k+2}{\lambda_1^4}\right)+(\Lambda-\lambda_1)\left(\frac{1}{(\Lambda-\lambda_1)^2}+\frac{2(N-k)-1}{(\Lambda-\lambda_1)^3}+\frac{3(N-k)^2-6(N-k)+2}{(\Lambda-\lambda_1)^4}\right)+o\left(\frac{1}{\Lambda^2}\right),\\
T_2 &= \frac{1}{\lambda_1}+\frac{k}{\lambda_1^2}+\frac{k(k-1)}{\lambda_1^3}+\frac{k(k-1)(k-2)}{\lambda_1^4}+ \frac{1}{\Lambda-\lambda_1}+\frac{N-k}{(\Lambda-\lambda_1)^2}+\frac{(N-k)(N-k-1)}{(\Lambda-\lambda_1)^3}+\frac{(N-k)(N-k-1)(N-k-2)}{(\Lambda-\lambda_1)^4} \nonumber \\ & -\lambda_1 \left(\frac{1}{\lambda_1^2}+\frac{2k}{\lambda_1^3}+\frac{3k(k-1)}{\lambda_1^4}+\frac{4k(k-1)(k-2)}{\lambda_1^4}\right)    - (\Lambda-\lambda_1)\left(\frac{1}{(\Lambda-\lambda_1)^2}+\frac{2(N-k)}{(\Lambda-\lambda_1)^3}+\frac{3(N-k)(N-k-1)}{(\Lambda-\lambda_1)^4} \right. \nonumber \\
& \left.+ \frac{4(N-k)(N-k-1)(N-k-2)}{(\Lambda-\lambda_1)^4}\right) +o\left(\frac{1}{\Lambda^3}\right). \nonumber 
\end{align*}}
Simplifying the above expression we get,
    \begin{eqnarray}
   \frac{d \lambda_1}{dk} =  \frac{-\left(\frac{1}{\lambda_1}+\frac{2k-1}{\lambda_1^2}+\frac{3k^2-6k+2}{\lambda_1^3}+\frac{1}{\Lambda-\lambda_1}+\frac{2(N-k)-1}{(\Lambda-\lambda_1)^2}+\frac{3(N-k)^2-6(N-k)+2}{(\Lambda-\lambda_1)^3}+o\left(\frac{1}{\Lambda^2}\right)\right)}{-\frac{k}{\lambda_1^2}-\frac{2k(k-1)}{\lambda_1^3}-\frac{3k(k-1)(k-2)}{\lambda_1^4}-\frac{N-k}{(\Lambda-\lambda_1)^2}-\frac{2(N-k)(N-k-1)}{(\Lambda-\lambda_1)^3}-\frac{3(N-k)(N-k-1)(N-k-2)}{(\Lambda-\lambda_1)^4}+o\left(\frac{1}{\Lambda^3}\right)}. \nonumber
\end{eqnarray}
Subtracting~$\lambda_1/k$ and by some simplification~$ \frac{d \lambda_1}{dk} - \frac{\lambda_1}{k}$ equals, 
 \begin{align*}
= & \frac{ \frac{k}{\Lambda-\lambda_1}-\frac{(N-k)\lambda_1}{(\Lambda-\lambda_1)^2}+T_3+o\left( \frac{1}{\Lambda^2} \right)
    }{k\left(\frac{k}{\lambda_1^2}+\frac{2k(k-1)}{\lambda_1^3}+\frac{3k(k-1)(k-2)}{\lambda_1^4}+\frac{N-k}{(\Lambda-\lambda_1)^2}+\frac{2(N-k)(N-k-1)}{(\Lambda-\lambda_1)^3}+\frac{3(N-k)(N-k-1)(N-k-2)}{(\Lambda-\lambda_1)^4}+o\left(\frac{1}{\Lambda^3}\right)\right)} \nonumber \\
    = & \frac{\frac{k(N-k)}{(\Lambda-\lambda_1)^2} \left(\frac{\Lambda-\lambda_1}{N-k}-\frac{\lambda_1}{k} \right) +T_3+o\left( \frac{1}{\Lambda^2} \right)
    }{k\left(\frac{k}{\lambda_1^2}+\frac{2k(k-1)}{\lambda_1^3}+\frac{3k(k-1)(k-2)}{\lambda_1^4}+\frac{N-k}{(\Lambda-\lambda_1)^2}+\frac{2(N-k)(N-k-1)}{(\Lambda-\lambda_1)^3}+\frac{3(N-k)(N-k-1)(N-k-2)}{(\Lambda-\lambda_1)^4}+o\left(\frac{1}{\Lambda^3}\right)\right)}. 
    \end{align*}
    where~$T_3 := \frac{k}{\lambda_1^2}+\frac{k\left[2(N-k)-1\right]}{(\Lambda-\lambda_1)^2}-\frac{2(N-k)(N-k-1)\lambda_1}{(\Lambda-\lambda_1)^3}+\frac{3k^2-4k}{\lambda_1^3}+\frac{k(3(N-k)^2-6(N-k)+2}{(\Lambda-\lambda_1)^3}-\frac{3(N-k)(N-k-1)(N-k-2)\lambda_1}{(\Lambda-\lambda_1)^4}.$
From Lemma~\ref{lem_accuracy_heavy}, it follows that~$\frac{\lambda_1}{\Lambda} = \frac{k}{N} + o(1)$ and~$\frac{\Lambda-\lambda_1}{\Lambda} = \frac{N-k}{N} + o(1)$ as~$\Lambda \to \infty,$ with the~$o(1)$ terms being uniform over~$k \in \left[N/2 ,N-\epsilon\right].$ Additionally, from Lemma~\ref{lem_accuracy_heavy_tight},~$\left(\frac{\lambda_1}{k} - \frac{\Lambda-\lambda_1}{N-k}\right) = \left(\frac{1}{N-k} - \frac{1}{k}\right) + o(1),$ with the~$o(1)$ term again being uniform over~$k \in \left[N/2,N-\epsilon\right].$ Now, multiplying  by~$\Lambda^2$ in the numerator and denominator above and applying these results, we obtain
    \begin{eqnarray}
   \lim_{\Lambda \to \infty} \left( \frac{d \lambda_1}{dk} - \frac{\lambda_1}{k}  \right) = \frac{\frac{N^2}{k}+\frac{kN^2}{(N-k)^2}-\frac{kN^2}{N-k}\left(\frac{1}{N-k}-\frac{1}{k} \right)}{kN^2\left(\frac{1}{k}+\frac{1}{N-k}\right)} = \frac{1}{k}  >  0. \nonumber 
\end{eqnarray}
Observe that the above limit is uniform over~$k \in \left[ \frac{N}{2},N-\epsilon \right]$.
\Halmos
\endproof

\begin{lemma}
\label{lem_accuracy_heavy}
     $\frac{\lambda_1}{\Lambda} \to \frac{k}{N} \text{ and } \frac{\Lambda-\lambda_1}{\Lambda} \to \frac{N-k}{N}$   
    $\text{ uniformly over } k \in \left[ \frac{N}{2}, N \right] \text{ as } \Lambda \to \infty.$
\end{lemma}
\proof{Proof:} 
We know that the blocking probability of a coalition with~$s$ servers and offered load~$a$, when~$\Lambda \to \infty$ is bounded as (see~\cite{harel}):~$
 1 - \frac{1}{\rho} < B(s,a) <  \frac{\rho}{1+\rho} \text{ where } \rho = a/s > 0. \nonumber
$
Using the upper bound for the larger coalition and the lower bound for the smaller coalition, the arrival rate~$\lambda_1$ at WE can be lower bounded by~$\hat{\lambda}_1,$ which satisfies:~$
    \frac{\frac{\hat{\lambda}_1}{k}}{1+\frac{\hat{\lambda}_1}{k}} = 1 - \frac{1}{\frac{\Lambda-\hat{\lambda}_1}{N-k}} 
  \implies  \hat{\lambda}_1 = k\left(\frac{\Lambda-N+k}{N}\right). $
Next, using the upper bound for the smaller coalition and the lower bound for the larger coalition, we obtain an upper bound~$\tilde{\lambda}_1$ of~$\lambda_1$ as follows:~$
     \nicefrac{\frac{\Lambda-\tilde{\lambda}_1}{N-k}}{\left (1+\frac{\Lambda-\tilde{\lambda}_1}{N-k} \right )} = 1 - \nicefrac{1}{\left(\nicefrac{\tilde{\lambda}_1}{k}\right)} 
  \implies   \tilde{\lambda}_1 = k \left(\frac{\Lambda+N-k}{N}\right). $
  
From the above, we obtain the following bounds on $\lambda_1$,~$
   k\left(\frac{\Lambda-N+k}{N}\right) \le \lambda_1 \le k\left(\frac{\Lambda+N-k}{N}\right). 
$
  It now follows that ~$
   \left | \frac{\lambda_1}{\Lambda} - \frac{k}{N} \right|  \stackrel{(a)}{\le} k\left(\frac{N-k}{N\Lambda} \right) \ \stackrel{(b)}{\le} \ \frac{N}{\Lambda}.
$
(the above inequalities lead to inequality $(a)$, while the  bound $(b)$ is obvious), which implies that~$
\lim_{\Lambda \to \infty} \left | \frac{\lambda_1}{\Lambda} - \frac{k}{N} \right| = 0 \text{ uniformly over } k \in \left[ \frac{N}{2},N \right].  
$
This implies the result.  \Halmos
\endproof

\begin{lemma}
\label{lem_error_taylor}
 Suppose~$f$ is~$m$-times differentiable on~$[0,\infty)$, such that~$f^{(m)}(t;k)$ is non-negative, monotonically increasing, and~$
    f^{(m)}(t;k) \le c_1 +c_2t^N  \text{ for all } t \ge 0$ and $k \in \left[ \frac{N}{2}, N - \epsilon \right]$,
 for some positive scalars~$c_1 \text{ and } c_2$. Further, $\nicefrac{\Lambda}{2} \le \lambda_1(k) \le \Lambda$ for all $k$.
Then
 \begin{eqnarray}
    \int_0^\infty f(t;k)e^{-\lambda_1(k) t} dt = \int_0^\infty \sum_{j=0}^{m-1} \left( f^{(j)}(0;k)\frac{t^j}{j!}\right)   e^{-\lambda_1(k) t} dt + o\left(\frac{1}{\Lambda^{m-1}}\right), \text{ with } f^{(0)}(\cdot;k) = f(\cdot;k), \nonumber 
\end{eqnarray}
 as~$\Lambda \to \infty.$ Here, the~$o\left(\frac{1}{\Lambda^{m-1}}\right) $  error is uniform over~$k \in \left[ \frac{N}{2}, N-\epsilon \right]$ for~$\epsilon >0$.
\end{lemma}

\proof{Proof:}
Using the Taylor expansion of~$f(t;k)$ (for any~$t$) around~$0$~\cite{rudin}, we have
 \begin{eqnarray}
    \int_0^\infty f(t;k)e^{-\lambda_1(k)t} dt = \int_0^\infty \sum_{j=0}^{m-1} \left( f^{(j)}(0;k)\frac{t^j}{j!}\right)e^{-\lambda_1(k) t} dt + \int_0^\infty   \frac{f^{(m)}(c(t);k) t^m}{m!}   e^{-\lambda_1(k) t} dt. \nonumber
\end{eqnarray}  for some~$c(t)$ strictly between~$0$ and~$t$. Observe that the residue term above can be upper bounded  as~$
    \int_0^\infty \frac{f^{(m)}(c(t);k) t^m}{m!} e^{-\lambda_1(k) t} dt \le \int_0^\infty \frac{f^{(m)}(t;k) t^m}{m!} e^{-\lambda_1(k)t } dt,
$
 since the~$m^{th}$ derivative of~$f(\cdot)$ is strictly monotonically increasing in~$t$ and as~$c(t) \le t$. Under the hypothesis of this lemma, we have an upper bound independent of $k \in \left[ \frac{N}{2}, N- \epsilon \right]$, which further can be upper bounded:
 \begin{eqnarray}
     \int_0^\infty \frac{f^{(m)}(t;k) t^m}{m!} e^{-\frac{\Lambda}{2} t} dt  \le  \int_0^\infty \frac{\left(c_1+c_2t^N\right) t^m}{m!} e^{-\frac{\Lambda}{2} t}  dt \stackrel{(a)}{=}   \ o\left( \frac{1}{\Lambda^{m-1}} \right), \nonumber
 \end{eqnarray}
 where equality~$(a)$ follows from simple calculations (involving the gamma function).
\Halmos
\endproof

 \begin{lemma}
\label{lem_accuracy_heavy_tight}
 For any $\epsilon>0$,~~$
     \frac{\lambda_1}{k}-\frac{\Lambda-\lambda_1}{N-k} \to \left(\frac{1}{N-k}-\frac{1}{k} \right) 
    \text{  uniformly over } k \in \left[ \frac{N}{2}, N-\epsilon \right] \text{ {\it as }} \Lambda \to \infty. $
\end{lemma}
\proof{Proof:}
Observe that~\eqref{eq_integral_pb} coincides with the WE equation for  integral values of~$k$ and~$N-k$. Relaxing~$k$ to be a real-valued number such that~$k \in \left[ \frac{N}{2},N-\epsilon\right]$, observe that each integral is of the form~$
\int_0^\infty f(t;k)e^{-\lambda_1t} dt\text{ or } \int_0^\infty f(t;k)e^{-(\Lambda-\lambda_1)t} dt.
$
In heavy traffic, since~$\lambda_1$ and~$\Lambda-\lambda_1$ tend to infinity (see Lemma~\ref{lem_accuracy_heavy}) the value of these integrals is dominated by the behavior of the integrand around zero. Accordingly, one can approximate these integrals using a Taylor expansion of~$f(t;k)$ around~$t=0.$ Formally, using Lemma~\ref{lem_error_taylor} (it is easy to show that all the above integrals satisfy the hypotheses of Lemma~\ref{lem_error_taylor}) and solving the non-negligible integrals, equation~\eqref{eq_integral_pb} can be re-written as~$
 1+ \frac{k}{\lambda_1}+\frac{k(k-1)}{\lambda_1^2} = 1+ \frac{N-k}{\Lambda-\lambda_1}+\frac{(N-k)(N-k-1)}{(\Lambda-\lambda_1)^2} + o\left(\frac{1}{\Lambda^2}\right), 
$ with~$o\left( \frac{1}{\Lambda^2} \right)$ being uniform over $k \in \left[ \frac{N}{2}, N-\epsilon \right]$.
 Simplifying the above using Lemma~\ref{lem_accuracy_heavy} (e.g., $o\left( \nicefrac{1}{\Lambda-\lambda_1}\right) = o\left( \nicefrac{1}{\Lambda}\right)$), and using $\Lambda o\left( \nicefrac{1}{\Lambda^2} \right) = o(\nicefrac{1}{\Lambda})$,  
 \begin{eqnarray}
\frac{k}{\lambda_1}\left( 1 + \frac{k-1}{\lambda_1} \right) & = & \frac{N-k}{\Lambda-\lambda_1}\left(1+\frac{N-k-1}{\Lambda-\lambda_1} + o\left( \frac{1}{\Lambda} \right)  \right) \implies
\frac{\lambda_1}{k} =
\frac{\Lambda-\lambda_1}{N-k} \left[\frac{\left( 1 + \frac{k-1}{\lambda_1} \right)}{\left(1+\frac{N-k-1}{\Lambda-\lambda_1} \right) + o\left( \frac{1}{\Lambda} \right)} \right].\nonumber
\end{eqnarray}
Subtracting~$\frac{\Lambda-\lambda_1}{N-k}$ from both sides of the above equation, we have
\begin{eqnarray}
\frac{\lambda_1}{k}-\frac{\Lambda-\lambda_1}{N-k} 
= \frac{\Lambda-\lambda_1}{N-k}\left[\frac{\left(  \frac{k-1}{\lambda_1} \right)-\left(\frac{N-k-1}{\Lambda-\lambda_1}\right)-o\left(\frac{1}{\Lambda}\right)}{\left(1+\frac{N-k-1}{\Lambda-\lambda_1} \right) + o\left( \frac{1}{\Lambda} \right)}\right]. \nonumber
\end{eqnarray}
Note that as~$\Lambda \to \infty$, the denominator of the above expression goes to~$1$. Further, multiplying and dividing by~$\Lambda$ and using Lemma~\ref{lem_accuracy_heavy}, we have (observe all errors converge uniformly in~$k$)
 \begin{eqnarray}
\lim_{\Lambda \to \infty} \left(\frac{\lambda_1}{k}-\frac{\Lambda-\lambda_1}{N-k} \right) = \frac{1}{N} \left[\left( \frac{k-1}{k}\right) N - \left( \frac{N-k-1}{N-k}\right)N\right] = \frac{1}{N-k} - \frac{1}{k}. \Halmos \nonumber 
\end{eqnarray}
\endproof

\begin{lemma}
While differentiating \eqref{eq_integral_pb}, the limits (derivative) and the integral can be interchanged.
\label{lem_interchange}
\end{lemma}
\proof{Proof:}
Since the blocking probability of any coalition increases with increase in arrival rate, the derivative of the left hand side of~\eqref{eq_integral_pb} with respect to~$\lambda_1(k)$ is not zero. Thus using Implicit Function Theorem, we obtain~$\lambda_1(k)$ to be a continuously differentiable function of~$k$ and hence,~$T_4 := \sup_{ k' \in [k, k+{\bar h}]} \frac{d \lambda_1(k')}{dk}$ is  finite for some~${\bar h}>0$.

It is sufficient to consider the limit of the form~$
    \lim_{h \to 0} \int_0^\infty \left( \frac{(1+t)^{k+h}e^{-\lambda_1(k+h) t}-(1+t)^ke^{-\lambda_1(k) t}}{h} \right)  dt.
$
By differentiability for all~$t$,

\vspace{-8mm}
{\small $$
 y_h(t) := \left( \frac{(1+t)^{k+h}e^{-\lambda_1(k+h) t}-(1+t)^ke^{-\lambda_1(k) t}}{h} \right)  \to \bigg((1+t)^k\ln(1+t) - (1+t)^kt(d\lambda_1(k)/dk)\bigg)e^{-\lambda_1(k) t}.
$$}
Consider any~$h \in (0, {\bar h} ]$. By Mean Value Theorem, there exists a~$k' \in (k,k+h)$ such that

\vspace{-9mm}
 {\small \begin{eqnarray}
 y_h(t) \le \left((1+t)^{k'}\ln(1+t)-(1+t)^{k'}t\left(\frac{d \lambda_1(k')}{dk}\right)\right)e^{-\lambda_1(k')t} \le \left((1+t)^{k+{\bar h}}\ln(1+t)+(1+t)^{k+{\bar h}}t T_4\right)e^{-\lambda_1(k)t}. \nonumber
\end{eqnarray}}
The upper bound is integrable and hence the result follows by Lebesgue's Dominated Convergence Theorem. \Halmos
 \endproof

 \proof{Proof of Theorem \ref{low_traffic}:} 
 $(i)$ Consider any~$2$-partition~$\P = \{C_1,C_2\} \in \mathscr{P}$ and~$\Phi \in {\bm \Phi}^\P$. From Theorem~\ref{Thm_two_partition}.$(iii)$,~$\P$ is stable under RB-IA rule.
  
Now, consider a~$2$-partition~$\P = \{C_1,C_2\} \notin \mathscr{P}$. This implies there exists a~$C \subset C_1$ such that~$N_{C_1} > N_{C} > N/2.$ We will show that coalition~$C$ blocks the configuration~$(\P,\Phi^\P_p)$. 
 From Lemma~\ref{lem_light_new}, we have~$
	\nicefrac{\lambda_{C_1}^\P}{\Lambda N_{C_1}} \to \nicefrac{1}{N_{C_1}} \text{ and } \nicefrac{\underline{\lambda}_C}{(\Lambda N_{C})} \to \nicefrac{1}{N_{C}} > \nicefrac{1}{N_{C_1}}  \mbox{ as } \Lambda \to 0.
$ Thus there exists a~$\underline{\Lambda}>0$ such that for any~$\Lambda \le \underline{\Lambda}$,~$
\nicefrac{\underline{\lambda}_C}{N_C} > \nicefrac{\lambda_{C_1}^\P}{N_{C_1}}.
 $
 It now follows that coalition~$C$ satisfies condition~\eqref{Eqn_condition_S} for blocking.  Moreover, under the proportional payoff vector~$\Phi_p^\P$,~\eqref{Eqn_condition_S} implies~\eqref{Eqn_condition_S_pt2}. This means that~$C$ blocks the configuration~$(\P,\Phi^\P_p)$, which in turn implies that~$\P$ is not a stable partition under RB-IA rule.
   
    $(ii)$ Under proportional payoff vector~$\Phi_p^\P$,~\eqref{eq_def_RBPA} is equivalent to~\eqref{Eqn_condition_S}, and hence the result under RB-PA follows along similar lines.
\Halmos
\endproof

\begin{lemma}
\label{lem_light_new}
Consider a coalition $C$ such that $N_C>N/2$, then~$
\frac{\underline{\lambda}_{C}}{\Lambda}
\to 1 \text{ as } \Lambda \to 0.
$
Consequently, for any~$2$-partition~$\P = \{C_1,C_2\}$ where~$N_{C_1} > N_{C_2}$,~$
\frac{\lambda_{C_1}^\P}{\Lambda} = \frac{\underline{\lambda}_{C_1}}{\Lambda}
\to 1 \text{ as } \Lambda \to 0.
$
\end{lemma}

\proof{Proof:}
Let~$\lambda_1 =  \underline{\lambda}_{C}.$  It is sufficient to show that~$
\frac{\lambda_1}{\Lambda-\lambda_1} \to \infty \text{ as } \Lambda \to 0.
$
In light traffic, the reciprocal of the blocking probabilities of the two coalitions satisfy
$$
R(k,\lambda_1) \sim \frac{k!}{\lambda_1^k} \text{ and } R(N-k,\Lambda-\lambda_1) \sim \frac{(N-k)!}{(\Lambda-\lambda_1)^{N-k}},
$$
where $f(\Lambda)\sim g(\Lambda)$ means $\lim_{\Lambda \to 0} \frac{f(\Lambda)}{g(\Lambda)} = 1.$ We therefore obtain,
\begin{eqnarray}
    \frac{k!}{\lambda_1^k}  \sim  \frac{(N-k)!}{(\Lambda-\lambda_1)^{N-k}} 
   \Rightarrow \left(\frac{\lambda_1}{\Lambda-\lambda_1}\right)^{N-k} \lambda_1^{2k-N}  \sim \frac{k!}{(N-k)!}. \nonumber
\end{eqnarray}
With~$\Lambda \to 0$,~$\lambda_1^{2k-N} \to 0$ and R.H.S. is a finite constant, this implies,~$
\lim_{\Lambda \to 0} \left(\frac{\lambda_1}{\Lambda-\lambda_1}\right) = \infty.
$
Now observe that for~$2$-partition~$\P = \{C_1,C_2\}$ with~$N_{C_1} > N_{C_2}$, we have~$N_{C_1} > N/2$ and hence the result follows.
\Halmos
\endproof

\end{APPENDICES}

%
%
%


%


\ACKNOWLEDGMENT{The first author's work is partially supported by the Prime Minister's Research Fellowship (PMRF), India. 
}



\begin{thebibliography}{}

\bibitem[Anily et al.(2010)]{anily2010}
Anily, S. and Haviv, M (2010) Cooperation in service systems. Operations Research, 58(3), pp.660-673.

\bibitem[Anily et al.(2011)]{anily2011}
Anily, S. and Haviv, M (2011) Homogeneous of degree one games are balanced with applications to service systems. Tel Aviv University, Faculty of Management, The Leon Recanati Graduate School of Business Administration.

\bibitem[Anily et al.(2014)]{anily2014}
Anily, S. and Haviv, M (2014) Subadditive and homogeneous of degree one games are totally balanced. Operations Research, 62(4), pp.788-793.

	\bibitem[{Aumann(1961)}]{aumann1961}
	Aumann, Robert J (1961)
The core of a cooperative game without side payments.
	Transactions of the American Mathematical Society, vol. 98, no. 3, pp. 539--552.

 \bibitem[{Aumann et al.(1974)}]{aumann1974cooperative}
	Aumann, Robert J and Dreze, Jacques H (1974)
  Cooperative games with coalition structures.
  International Journal of game theory,
  vol. 3, no. 4, pp. 217--237.
  
  \bibitem[Bloch, F.(1996)]{bloch}
  Bloch, F (1996) Sequential formation of coalitions in games with externalities and fixed payoff division. Games and economic behavior, 14(1), pp.90-123.

 	\bibitem[{Bloch et al.(2014)}]{pessimistic}
	Bloch, Francis and Van den Nouweland, Anne (2014) Expectation formation rules and the core of partition function games.
	Games and Economic Behavior, vol. 88, pp. 339--353.

 	\bibitem[{Correa et al.(2010)}]{WE}
	Correa, Jos{\'e} R and Stier-Moses, Nicol{\'a}s E (2010)
	Wardrop equilibria.
	Wiley encyclopedia of operations research and management science.
	
	\bibitem[García-Sanz et al.(2008)]{garcia}
	García-Sanz, M.D., Fernández, F.R., Fiestras-Janeiro, M.G., García-Jurado, I. and Puerto, J (2008) Cooperation in Markovian queueing models. European Journal of Operational Research, 188(2), pp.485-495.
	
	\bibitem[González, P. et al.(2004)]{gonzalez}
	González, P. and Herrero, C (2004) Optimal sharing of surgical costs in the presence of queues. Mathematical Methods of Operations Research, 59, pp.435-446.

 \bibitem[Hafalir(2007)]{hafalir}
Hafalir, Isa E (2007) Efficiency in coalition games with externalities. Games and Economic Behavior, 61(2), pp.242-258.

 
\bibitem[{Hajdukov{\'a}, Jana(2006)}]{CFGs}
	Hajdukov{\'a}, Jana (2006) Coalition formation games: A survey.
	International Game Theory Review, vol. 8, no. 04, pp. 613--641.

  \bibitem[{Harel(1988)}]{harel}
 Harel, A (1988) Sharp bounds and simple approximations for the Erlang delay and loss formulas. Management Science, 34(8), pp. 959-972.

 \bibitem[{Jagerman(1974)}]{jagerman}
 Jagerman, D.L (1974) Some properties of the Erlang loss function. Bell System Technical Journal, 53(3), pp.525-551.
 
 


\bibitem[{Karsten et al.(2012)}]{karsten2012}
Karsten, F., Slikker, M. and van Houtum, G.J (2012) Inventory pooling games for expensive, low‐demand spare parts. Naval Research Logistics (NRL), 59(5), pp.311-324.

\bibitem[{Karsten et al.(2014)}]{karsten2014}
Karsten, F., Slikker, M. and van Houtum, G.J (2014) Domain extensions of the Erlang loss function: Their scalability and its applications to cooperative games. Probability in the Engineering and Informational Sciences, 28(4), pp.473-488.

	\bibitem[{Karsten et al.(2015)}]{karsten}
 Karsten, Frank and Slikker, Marco and Van Houtum, Geert-Jan (2015)
Resource pooling and cost allocation among
independent service providers.
Operations Research, vol. 63, no. 2, pp. 476--488, INFORMS.

	\bibitem[{(Martins-da-Rocha et al.(2011))}]{alpha-core}
	Martins-da-Rocha, Victor Filipe and Yannelis, Nicholas C (2011)
	Non-emptiness of the alpha-core.
	Funda{\c{c}}{\~a}o Getulio Vargas. Escola de P{\'o}s-gradua{\c{c}}{\~a}o em Economia.

 \bibitem[{Narahari(2014)}]{narahari}
Narahari, Y (2014) Game theory and mechanism design. vol. 4. (World Scientific).

\bibitem[Özen et al.(2011)]{ozen}
Özen, U., Reiman, M.I. and Wang, Q (2011) On the core of cooperative queueing games. Operations Research Letters, 39(5), pp.385-389.

\bibitem[Ray, D. et al.(1999)]{ray}
Ray, D. and Vohra, R (1999) A theory of endogenous coalition structures. Games and economic behavior, 26(2), pp.286-336.


	\bibitem[{Rudin(1976))}]{rudin}
	Rudin, W (1976) Principles of mathematical analysis (vol. 3) (New York: McGraw-hill).
	
	\bibitem[{Saad, W. et al.(2011)}]{saad_unilateral}
	Saad, W., Han, Z., Zheng, R., Hjorungnes, A., Basar, T. and Poor, H.V (2011) Coalitional games in partition form for joint spectrum sensing and access in cognitive radio networks. IEEE Journal of Selected Topics in Signal Processing, 6(2), pp.195-209.
	
	
	\bibitem[{Saad, W., et al.(2009)}]{saad}
	Saad, W., Han, Z., Debbah, M., Hjorungnes, A. and Basar, T (2009) Coalitional game theory for communication networks. Ieee signal processing magazine, 26(5), pp.77-97.
	
	

	\bibitem[(Shiksha et al.(2021))]{Shiksha_Perf}
	Shiksha Singhal and Veeraruna Kavitha (2021)
Coalition Formation Resource Sharing Games in Networks.
	Performance Evaluation,
vol. 152,
102239,
ISSN 0166-5316.


\bibitem[{Singhal et al.(2021)}]{CDC} Singhal, S., Kavitha, V. and Nair, J (2021) Coalition formation in constant sum queueing games. In 2021 60th IEEE Conference on Decision and Control (CDC) (pp. 3812-3817). IEEE.

\bibitem[Thrall et al.(1963)]{lucas}
Thrall, R.M. and Lucas, W.F (1963) N‐person games in partition function form. Naval Research Logistics Quarterly, 10(1), pp.281-298.

\bibitem[Timmer et al.(2010)]{timmer}
Timmer, J. and Scheinhardt, W (2010) How to share the cost of cooperating queues in a tandem network?. In 2010 22nd International Teletraffic Congress (ITC 22) (pp. 1-7). IEEE.

\bibitem[Yi, S.S.( 1997)]{yi}
Yi, S.S (1997) Stable coalition structures with externalities. Games and economic behavior, 20(2), pp.201-237.


\bibitem[Yu et al.(2015)]{yu}
Yu, Y., Benjaafar, S. and Gerchak, Y (2015) Capacity sharing and cost allocation among independent firms with congestion. Production and Operations Management, 24(8), pp.1285-1310.

	
	
	





\end{thebibliography}



\end{document}